\documentclass[apj]{emulateapj}

\def\ra#1#2#3{#1$^{\rm h}$#2$^{\rm m}$#3$^{\rm s}$}
\def\dec#1#2#3{$#1^\circ#2'#3''$}
\def\lesssim{\mathrel{\hbox{\rlap{\hbox{\lower4pt\hbox{$\sim$}}}\hbox{$<$}}}}
\def\gtrsim{\mathrel{\hbox{\rlap{\hbox{\lower4pt\hbox{$\sim$}}}\hbox{$>$}}}}

\usepackage{epsfig,graphicx}
\usepackage{amsmath}
\usepackage{comment}
\usepackage{url}
\usepackage[normalem]{ulem}

\begin{document}

\title{\lowercase{i}PTF\,16asu: A Luminous, Rapidly-Evolving, and High-Velocity Supernova}
\email{lwhitesides@caltech.edu}

\def\cit{1}
\def\okc{2}
\def\ljmu{3}
\def\ttu{4}
\def\ngsfc{5}
\def\jsi{6}
\def\jpl{7}
\def\konus{8}
\def\ssl{9}
\def\wis{10}
\def\dark{11}
\def\ipac{12}
\def\lbnl{13}
\def\ucb{14}
\def\pu{15}
\def\isdc{16}
\def\lanl{17}

\author{L.~Whitesides\altaffilmark{\cit}, 
 R.~Lunnan\altaffilmark{\okc,\cit},
 M.~ M.~Kasliwal\altaffilmark{\cit},
 D.~A.~Perley\altaffilmark{\ljmu},
 A.~Corsi\altaffilmark{\ttu},
 S.~B.~Cenko\altaffilmark{\ngsfc,\jsi}, 
 N.~Blagorodnova\altaffilmark{\cit},
 Y.~Cao\altaffilmark{\cit}, 
 D.~O.~Cook\altaffilmark{\cit},
 G.~B.~Doran\altaffilmark{\jpl},
 D.~D.~Frederiks\altaffilmark{\konus}, 
 C.~Fremling\altaffilmark{\okc},
 K.~Hurley\altaffilmark{\ssl},
 E.~Karamehmetoglu\altaffilmark{\okc},  
 S.~R.~Kulkarni\altaffilmark{\cit},
 G.~Leloudas\altaffilmark{\wis,\dark},
 F.~Masci\altaffilmark{\ipac}, 
 P.~E.~Nugent\altaffilmark{\lbnl,\ucb}, 
 A.~Ritter\altaffilmark{\pu}, 
 A.~Rubin\altaffilmark{\wis},
 V.~Savchenko\altaffilmark{\isdc},
 J.~Sollerman\altaffilmark{\okc},
 D.~S.~Svinkin\altaffilmark{\konus}, 
 F.~Taddia\altaffilmark{\okc}, 
 P.~Vreeswijk\altaffilmark{\wis}, and
 P.~Wozniak\altaffilmark{\lanl} 
 }

\altaffiltext{\cit}{Department of Astronomy, California Institute of Technology, 1200 East California Boulevard, Pasadena, CA 91125, USA}
\altaffiltext{\okc}{The Oskar Klein Centre \& Department of Astronomy, Stockholm University, AlbaNova, SE-106 91 Stockholm, Sweden}
\altaffiltext{\ljmu}{Astrophysics Research Institute, Liverpool John Moores University, IC2, Liverpool Science Park, 146 Browlow Hill, Liverpool L3 5RF, UK}
\altaffiltext{\ttu}{Department of Physics and Astronomy, Texas Tech University, Box 41051, Lubbock, TX 79409-1051, USA}
\altaffiltext{\ngsfc}{Astrophysics Science Division, NASA Goddard Space Flight Center, Mail Code 661, Greenbelt, MD 20771, USA}
\altaffiltext{\jsi}{Joint Space-Science Institute, University of Maryland, College Park, MD 20742, USA}
\altaffiltext{\jpl}{Jet Propulsion Laboratory, California Institute of Technology, Pasadena, CA 91109, USA}
\altaffiltext{\konus}{Ioffe Institute, Politekhnicheskaya 26, St. Petersburg 194021, Russia}
\altaffiltext{\ssl}{University of California,
Berkeley, Space Sciences Laboratory, 7 Gauss Way, Berkeley, CA
94720-7450, USA}
\altaffiltext{\wis}{Department of Particle Physics and Astrophysics, Weizmann Institute of Science, 234 Herzl St., Rehovot, Israel}
\altaffiltext{\dark}{Dark Cosmology Centre, Niels Bohr Institute, University of Copenhagen, Juliane Maries Vej 30, DK-2100 Copenhagen, Denmark}
\altaffiltext{\ipac}{Infrared Processing and Analysis Center, California Institute of Technology, MS 100-22, Pasadena, CA 91125, USA}
\altaffiltext{\lbnl}{Lawrence Berkeley National Laboratory, Berkeley, California 94720, USA}
\altaffiltext{\ucb}{Department of Astronomy, University of California, 501 Campbell Hall, Berkeley, CA 94720-3411, USA}
\altaffiltext{\pu}{Department of Astrophysical Sciences, Princeton University, Princeton, NJ 08544, USA}
\altaffiltext{\isdc}{ISDC, Department of astronomy, University of Geneva, chemin d’Écogia, 16 CH-1290 Versoix, Switzerland}
\altaffiltext{\lanl}{Los Alamos National Laboratory, MS-D466, Los Alamos, NM 87545, USA}

\begin{abstract}
Wide-field surveys are discovering a growing number of rare transients whose physical origin is not yet well understood. Here, we present optical and UV data and analysis of iPTF\,16asu, a luminous, rapidly-evolving, high velocity, stripped-envelope supernova. With a rest-frame rise-time of just 4~days and a peak absolute magnitude of $M_{\rm g}=-20.4$~mag, the light curve of iPTF\,16asu is faster and more luminous than previous rapid transients. The spectra of iPTF\,16asu show a featureless, blue continuum near peak that develops into a Type Ic-BL spectrum on the decline. We show that while the late-time light curve could plausibly be powered by $^{56}$Ni decay, the early emission requires a different energy source. Non-detections in the X-ray and radio strongly constrain the energy coupled to relativistic ejecta to be at most comparable to the class of low-luminosity gamma-ray bursts. We suggest that the early emission may have been powered by either a rapidly spinning-down magnetar, or by shock breakout in an extended envelope of a very energetic explosion. In either scenario a central engine is required, making iPTF\,16asu an intriguing transition object between superluminous supernovae, Type Ic-BL supernovae, and low-luminosity gamma-ray bursts. 
\end{abstract}

\keywords{supernovae: general; supernovae: individual (iPTF16asu); gamma-ray burst: general; magnetars; shock waves}

\section{Introduction}
Many new and unusual astrophysical transients have been discovered recently by wide-field surveys which regularly monitor the night sky. Supernovae (SNe) are traditionally classified based on their spectra (see \citealt{fil97} for a review) and fall into two main groups: Type II/Ibc SNe, which originate from core collapse of massive stars; and Type Ia SNe, which are produced by thermonuclear disruptions of white dwarfs. The advent of dedicated wide-field surveys with increased survey speeds has lead to the discovery of exotic types of SNe and other transient events both inside and outside of galaxies (see \citealt{k12} for a review). These rare detections have necessitated the establishment of new categories of SNe such as Ca-rich gap transients (e.g., \citealt{pgm+10}), .Ia explosions (e.g., \citealt{kkg+10}), Intermediate Luminosity Red Transients (e.g., \citealt{pkt+08}), and superluminous supernovae (e.g., \citealt{qkk+11}) which demand different physical models than those previously used to explain SNe. The physics powering transient objects in our universe continues to be a rich topic of exploration.

\begin{figure}
\centering
\includegraphics[width=3.5in]{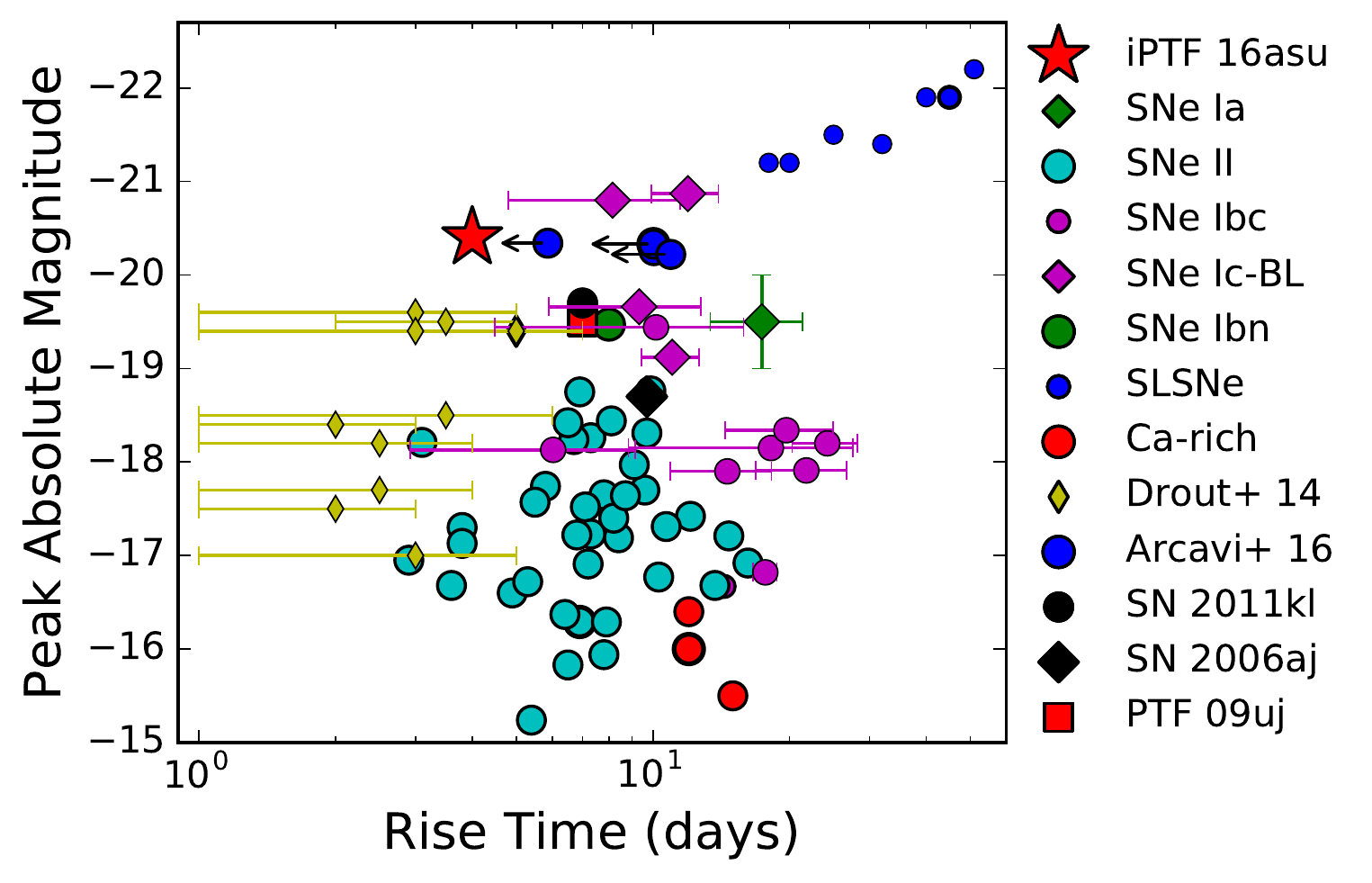}
\caption{Rest frame rise time (explosion to peak) versus peak absolute magnitude of a variety of types of SNe. iPTF\,16asu, shown as a red star, is unique in its combination of high luminosity and fast rise time. Data from \citet{hgk+10} ($B$~band), \citet{rgd+16} ($R$~band), \citet{tsl+15} ($g$~band), \citet{hav+17} (template), \citet{bdt+09} ($i$~band), \citet{psb+10} ($B$~band), \citet{qkk+11} ($u$~band), \citet{lcb+13} ($r$~band), \citet{isj+13} ($r$~band), \citet{ccs+11} ($z$~band), \citet{dcs+14} ($r$~band), \citet{awh+16} ($r$~band), \citet{gmk+15} ($i$~band), \citet{lkc+17} ($r$~band), \citet{kkg+12} ($r$~band), and \citet{orn+10} (NUV~band). Where possible rise times are given in band closest to iPTF\,16asu rest-frame $g$~band. 
\label{fig:intro}}
\end{figure}

This diverse landscape of transients is illustrated in Figure~\ref{fig:intro}, shown in the phase space of rise time (explosion to peak) versus peak luminosity. As peak luminosities are not always available in the same filters, this figure should not be used for quantitative comparisons, but rather as an illustration of the approximate areas inhabited by different transients in this phase space. Type Ia SNe, shown as a green diamond, act as standardizable candles \citep{phi93}. They exhibit a tight range of luminosities and rise times \citep{hgk+10}. Type II SNe, shown in cyan, are characterized by fast rise times but relatively low luminosities \citep{rgd+16}. Type Ibc SNe, shown in magenta, are more heterogeneous but tend to rise more slowly and become brighter than Type II \citep{tsl+15}; those with broad spectral features (Type Ic-BL), denoted as diamonds, generally reach higher peak luminosities than typical SNe Ibc \citep{cog+12, cck+17}. Superluminous supernovae (SLSNe), shown in blue, are extremely bright transients with very long rise times \citep{qkk+11, gal12}.

Transients which rise and decay rapidly are difficult to detect, requiring a sufficiently high cadence over a sufficiently large volume, rapid triggering and follow-up. Improvements in these areas have enabled discovery of objects which populate this previously empty region of short time scales at a wide range of luminosities. \citet{dcs+14} searched the Pan-STARRS Medium Deep Survey for rapidly evolving transients, resulting in the sample of objects shown in yellow in Figure~\ref{fig:intro}. Recently, \citet{awh+16} presented another four rapidly-evolving objects (shown in blue), with intermediate luminosities between SLSNe and the other types of known SNe. These objects are also similar in rise time and luminosity to SN\,2011kl, a unique event which was associated with an ultra-long gamma-ray burst (GRB), shown in black \citep{gmk+15, kso+16}. Interaction-powered SNe, including Type Ibn SNe, can also show short rise times and high peak luminosities (e.g., \citealt{hav+17,orn+10}).

Here we present an analysis of iPTF\,16asu, a transient with peak magnitude intermediate between SLSNe and ordinary SNe ($M_{\rm g}=-20.4~{\rm mag}$) and an extremely rapid ($4.0$ day) rise to peak; shown as a red star in Figure~\ref{fig:intro}. These characteristics place iPTF\,16asu in a neighboring, but unique part of transient phase space to the objects analyzed in \citet{awh+16} and \cite{dcs+14}. We present the photometric and spectroscopic observations of iPTF\,16asu in Section~\ref{sec:data}; analyze the light curve and spectra in Section~\ref{sec:lc} and Section~\ref{sec:spec_analysis}; and discuss the host galaxy in Section~\ref{sec:hostgal}. We discuss the feasibility of several physical explosion mechanisms and energy sources for iPTF\,16asu in Section~\ref{sec:models}, and summarize our findings in Section~\ref{sec:conc}.

Throughout this paper, we assume a flat $\Lambda$CDM cosmology with $\Omega_{\rm M}=0.286$ and H$_{0}=69.6~{\rm km~s}^{-1}{\rm Mpc}^{-1}$.

\section{Observations}
\label{sec:data}

\subsection{Intermediate Palomar Transient Factory Discovery}
iPTF\,16asu was discovered by the intermediate Palomar Transient Factory (iPTF; \citealt{lkd+09}, \citealt{cnk16}, \citealt{mlr+17}), and was first detected in data taken with the 48-inch Samuel Oschin Telescope at Palomar Observatory (P48) on 2016 May 11.26 UT (UT dates are used throughout this paper) at coordinates RA=\ra{12}{59}{09.28}, Dec=\dec{+13}{48}{09.2} (J2000.0) and at a magnitude of $g=20.54$~mag. We obtained a spectrum with the Double Beam Spectrograph (DBSP; \citealt{og82}) on the 200-inch Hale Telescope at Palomar Observatory on 2016 May 14.3, which shows a blue continuum and narrow H$\alpha$ and [\ion{O}{3}] lines from the host galaxy, setting the redshift to $z=0.187$. A later spectrum taken on 2016 June 04 by the DEep Imaging Multi-Object Spectrograph (DEIMOS; \citealt{fpk+03}) on the 10-m Keck II telescope on 2016 Jun 04 shows SN features consistent with a Type Ic-BL SN. The spectroscopic evolution is discussed in Section~\ref{sec:spec_analysis}.

\subsection{Photometry}
\begin{figure}
\centering
\includegraphics[width=3.5in]{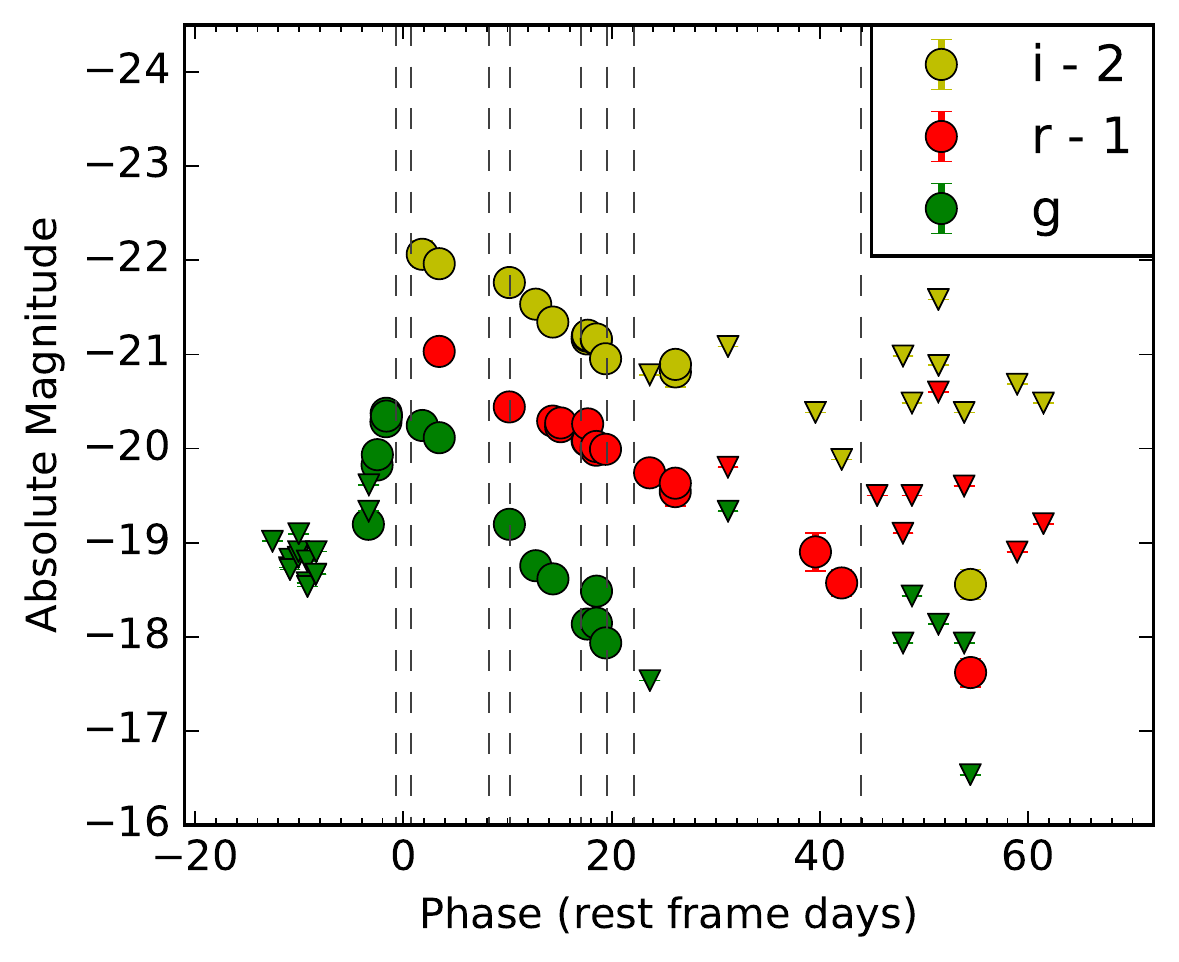}
\caption{Light curves of iPTF\,16asu in $g$, $r$, and $i$ filters. As indicated in the legend, the $r$ and $i$~band data have been offset for clarity. Triangles denote non-detections. Dashed lines indicate times of spectroscopic observations.
\label{fig:16asu_lc}}
\end{figure}

iPTF\,16asu was detected in a nightly-cadence $g$~band experiment with iPTF, and we therefore have P48 data covering the time up to explosion as well as the early rise. Subsequent photometry was obtained with the automated 60-inch telescope at Palomar (P60; \citealt{cfm+06}) in the $gri$ bands. Host-subtracted point-spread function (PSF) photometry was obtained using the Palomar Transient Factory Image Differencing and Extraction (PTFIDE) pipeline \citep{mlr+17} on the P48 images, and the FPipe SEDM presented in \citet{fst+16} on the P60 images using Sloan Digital Sky Survey (SDSS; \citealt{sdss13}) images as templates, and also calibrating to SDSS. Our last photometric observation came from the 3.58-m Telescopio Nazionale Galileo (TNG) and was processed through the FPipe. The photometry has been corrected for Galactic extinction following \citet{sf11}, with E${(B-V)} = 0.029$~mag and all magnitudes in this paper are reported in the AB system. Table~\ref{tab:phot} lists all photometric data, which is shown in Figure~\ref{fig:16asu_lc}.

\subsection{Spectroscopy}
\begin{figure}
\centering
\includegraphics[width=3.5in]{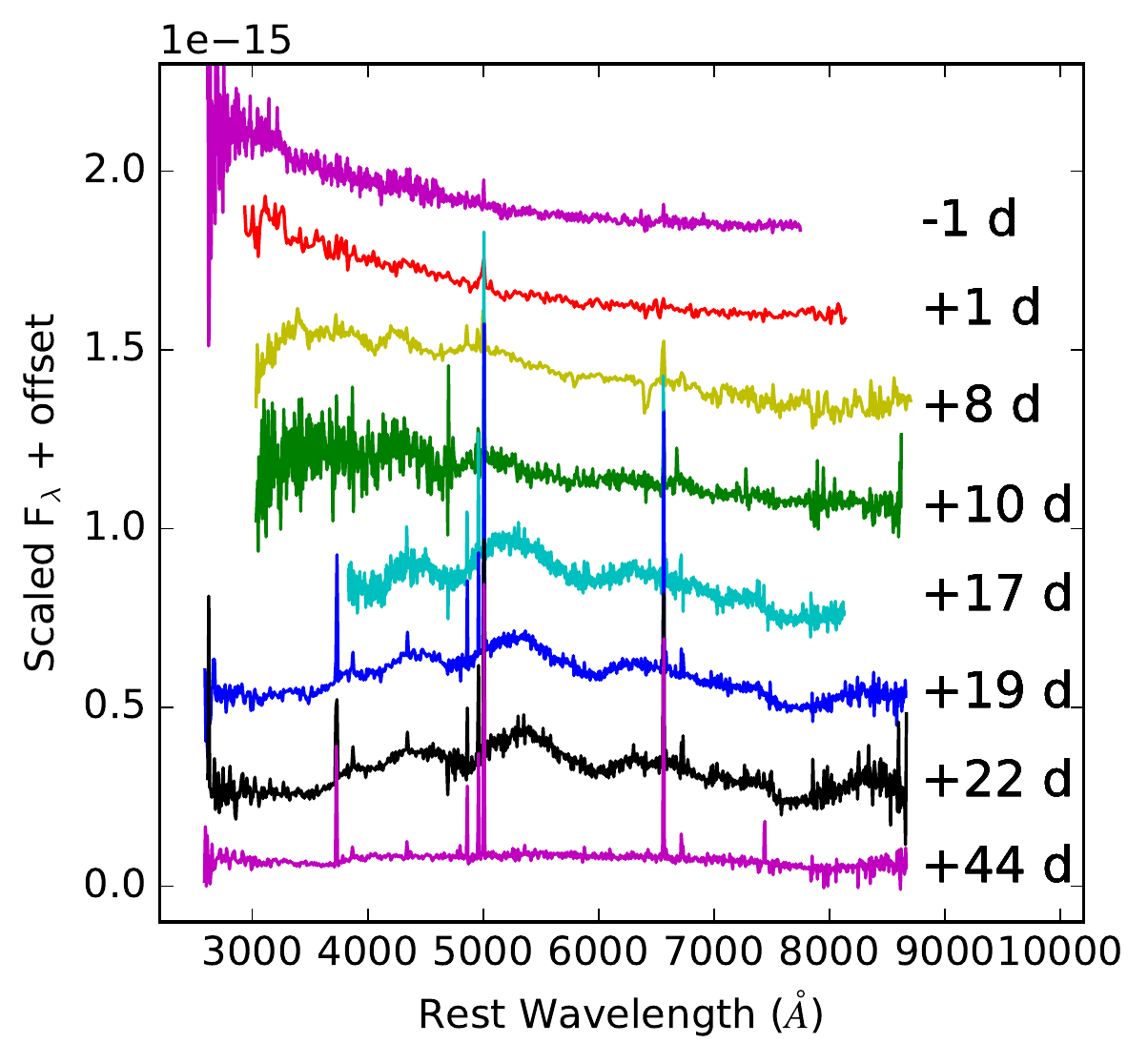}
\caption{Sequence of observed spectra for iPTF\,16asu. Phase in rest-frame days relative to $g$~band maximum is given to the right of each spectrum. The first two spectra show a featureless, blue continuum, with broad SN features starting to be visible 8 days after maximum. By day 17, the spectrum has developed into that of a Type Ic-BL SN. Our last spectrum, taken 44 days after peak, is dominated by galaxy light. Galaxy narrow emission lines have not been removed. Spectra have been binned and arbitrarily scaled for display purposes. See Section~\ref{sec:spec_analysis} for details.
\label{fig:spectra}}
\end{figure}

We obtained a sequence of eight low resolution spectra for iPTF\,16asu using the DBSP on P200; the Andalucia Faint Object Spectrograph and Camera (ALFOSC) on the 2.56-m Nordic Optical Telescope (NOT); the Device Optimized for the LOw RESolution (DOLORES) on TNG; the Low Resolution Imaging Spectrometer (LRIS; \citealt{occ+95}) on Keck I, and the DEIMOS on the 10-m Keck II telescope. The times of the spectra are marked as dashed lines in Figure~\ref{fig:16asu_lc}, and details of the spectroscopic observations are given in Table~\ref{tab:spec}. 
Spectra were reduced using standard procedures using IRAF\footnote{IRAF is distributed by the National Optical Astronomy Observatory, which is operated by the Association of Universities for Research in Astronomy (AURA) under a cooperative agreement with the National Science Foundation.} and IDL, including wavelength calibration using arc lamps, and flux calibration using standard stars. The spectroscopic sequences for iPTF\,16asu is shown in Figure~\ref{fig:spectra}, and the spectroscopic properties are analyzed and discussed in Section~\ref{sec:spec_analysis}. All spectra will be made available in the Weizmann Interactive Supernova Data Repository (WISeREP; \citealt{wiserep}).

\subsection{Radio Observations}
We observed the field of iPTF\,16asu with the Karl G. Jansky Very Large Array (VLA) on two epochs (Program VLA/16B-043; PI: A.~Corsi). The first observation was carried out starting on 2016 June 13, 01:18:22 UT (MJD 57552), with the VLA in its B configuration. The second observation was carried out with the VLA in its A configuration, starting on 2017 January 10, 09:43:06 UT (MJD 57763). Both these observations were carried out in C-band (nominal central frequency of $\approx 5$~GHz), using the 8 bit configuration and 2 GHz nominal bandwidth. On both epochs we used 3C286 as bandpass and flux density calibrator, and J1300+1417 as phase calibrator. The total observing time was about 1~hr (including calibration and overhead) per epoch.

VLA data were calibrated using the automated VLA calibration pipeline in CASA \citep{McMullin2007}. After visual inspection, additional flags were applied when needed. Images of the fields were produced using the CLEAN task \citep{Hogbom1974}.

We searched for a radio counterpart to iPTF\,16asu within a 2''-radius circle centered on the iPTF position of iPTF\,16asu. No radio source was detected within this region down to a 3$\sigma$ limit of $\approx 17\,\mu$Jy at 6.2\,GHz for both epochs.

\subsection{UV and X-Ray Observations}
At the time of the first spectrum, iPTF\,16asu resembled a very young SLSN, with its already high luminosity and blue spectrum indicating a high temperature. We therefore triggered our \textit{Swift} program for SLSNe (GI-1215281, PI: R.~Lunnan), and three epochs of \textit{Swift} UVOT \citep{uvot} and XRT \citep{xrt} data were obtained, at phases corresponding to 7.4, 13.4 and 19.2 days after explosion (see Section~\ref{sec:lc_analysis} for calculation of explosion date). 

We reduced the \textit{Swift} data using the HEASoft package provided by NASA\footnote{\url http://heasarc.nasa.gov/lheasoft/}. UVOT photometry was performed using the task {\tt UVOTsource} with an aperture of 5\arcsec. iPTF\,16asu is detected in all filters except $V$~band in the first observation, and undetected in all UVOT filters in the subsequent two epochs, due to the rapid fading of the SN. All UVOT photometry is listed in Table~\ref{tab:phot}.

The XRT data were reduced with the {\tt Ximage} software from the HEASoft package. No X-ray source is detected at the position of iPTF\,16asu in either epoch. The $3\sigma$ upper limits correspond to $5.6\times10^{-3}$~counts~s$^{-1}$, $2.9\times10^{-3}$~counts~s$^{-1}$ and $3.9\times10^{-3}$~counts~s$^{-1}$, respectively. Using WebPIMMS\footnote{\url http://heasarc.gsfc.nasa.gov/cgi-bin/Tools/w3pimms/w3pimms.pl} and assuming a Galactic nH of $2.2\times10^{20}{\rm cm}^{-2}$, we find that $1\times 10^{-3}$~counts~s$^{-1}$ corresponds to $3.76\times10^{-14}~{\rm erg~cm}^{-2}~{\rm s}^{-1}$ (unabsorbed; 0.3-10~keV), assuming a power law model with an index of 2. At a redshift of $z=0.1874$, our X-ray count limits translates to flux limits of $2.5\times10^{43}~{\rm erg~s}^{-1}$, $1.1\times10^{43}~{\rm erg~s}^{-1}$ and $1.5\times10^{43}~{\rm erg~s}^{-1}$ respectively.

\subsection{Search for Associated Gamma-Ray Burst}
\label{sec:konus}
We searched the Gamma-Ray Coordinates Network (GCN) archives for any announced GRBs consistent with the location and best-fit explosion time of iPTF\,16asu (Section~\ref{sec:lc_analysis}). No announced GRB was consistent with the location and time of iPTF\,16asu, also when extending the search to include bursts detected between the last iPTF non-detection and the first detection of iPTF\,16asu. However, our analysis of the Konus-Wind data (KW; \citealt{konus}) reveals that a weak burst was detected by KW in the waiting mode (with time resolution of 2.944~s) on 2016 May 10.41, which is consistent with our best-fit explosion time of 2016 May $10.53 \pm 0.17$ days (see Section~\ref{sec:lc_analysis}). The burst was observed by the KW S2 detector pointing the northern ecliptic hemisphere (nothing is seen in the opposite S1 detector), which is also consistent with the position of iPTF\,16asu, but the burst source position cannot be constrained more precisely from the KW data. 

The burst emission is significant in two softest KW energy bands: G1 (20-80~keV, $13\sigma$) and G2 (80-300~keV, $8\sigma$). The burst light curve shows a single emission episode with a duration of 126 s (T$_{50}$ = $56 \pm 11$~s and T$_{90}$ = $100 \pm 11$~s, both measured in the 20-300~keV energy band). Fitting the KW tree-channel time-integrated spectrum (measured from T0 to T0+126.592~s) by a simple power law yields the photon index of 2.35$_{-0.14}^{+0.18}$, $\chi^ 2$/dof = 2.7/1. From this fit, the burst had an energy fluence of 8.25$_{-0.86}^{+1.60} \times 10^{-6} {\rm erg~cm}^{-2}$ and a 2.944-s peak energy flux, measured from T0+73.6~s, of $2.41_{-0.94}^{+1.02} \times 10^{-7}~{\rm erg~cm}^{-2}{\rm s}^{-1}$ (both in the 20-1200~keV energy range). At the distance of iPTF\,16asu, this fluence would correspond to an equivalent isotropic energy $E_{\rm iso}$ of $8.2\times 10^{50}~{\rm erg}$. The fit with a power law with exponential cutoff model yields only an upper limit on spectrum peak energy: $E_{\rm p} < 67~{\rm keV}$.

During the KW burst, \textit{Swift} was in SAA and the position of iPTF\,16asu was Earth-occulted. However, the position of iPTF\,16asu was not occulted for \textit{Fermi} (and six GBM detectors had incident angles less than 60 deg). We analyze the \textit{Fermi}-GBM continuous data, and find no emission in the 30-300 keV band coincident with KW burst. Given that the background of \textit{Fermi}-GBM is considerably lower than KW, this implies that the KW burst came from a source Earth-occulted to \textit{Fermi}, and therefore is not related to iPTF\,16asu. 

We also searched for a possible GRB in the INTEGRAL-SPI-ACS (SPI-ACS; \citealt{vbr+03}) data covering the 75--8000~keV range and found no candidate event down to the 3 sigma level.
Since KW and SPI-ACS were observing the whole sky during the interval of interest, upper limits on gamma-ray flux can be obtained. For the whole interval (excluding the KW burst), assuming a typical long GRB spectrum (the Band function with $\alpha=-1$, $\beta=-2.5$, and $E_{\rm p}=300~{\rm keV}$), the corresponding KW and SPI-ACS limiting peak fluxes estimates are $\sim (1-4) \times 10^{-7}~{\rm erg~cm}^{-2}~{\rm s}^{-1}$, both in the 10~keV - 10~MeV band at 3--10~s time scales.

We conclude therefore that there is no statistically significant evidence for a SN-associated GRB down to threshold of $10^{-7}~{\rm erg~cm}^{-2}{\rm s}^{-1}$. The associated isotropic peak luminosity limit is $L_{\rm iso} \lesssim 10^{49} {\rm erg~s}^{-1}$ and total energy $E_{\rm iso} \lesssim 10^{50}~{\rm erg}$ (both calculated in the 10~keV - 10~MeV energy range). Hence, from these limits, an accompanying low-luminosity GRB, like GRB 980425 ($L_{\rm iso} \sim 5 \times 10^{46}~{\rm erg~s}^{-1}$, $E_{\rm iso} \sim 10^{48}~{\rm erg}$; \citealt{gvv+98}), cannot be excluded.
We return to discuss possible GRB models for iPTF\,16asu in Section~\ref{sec:grb}.

\section{Light Curve Analysis}
\label{sec:lc}

\subsection{Rise Time and Peak Luminosity}
\label{sec:lc_analysis}

\begin{figure}
\centering
\includegraphics[width=3.5in]{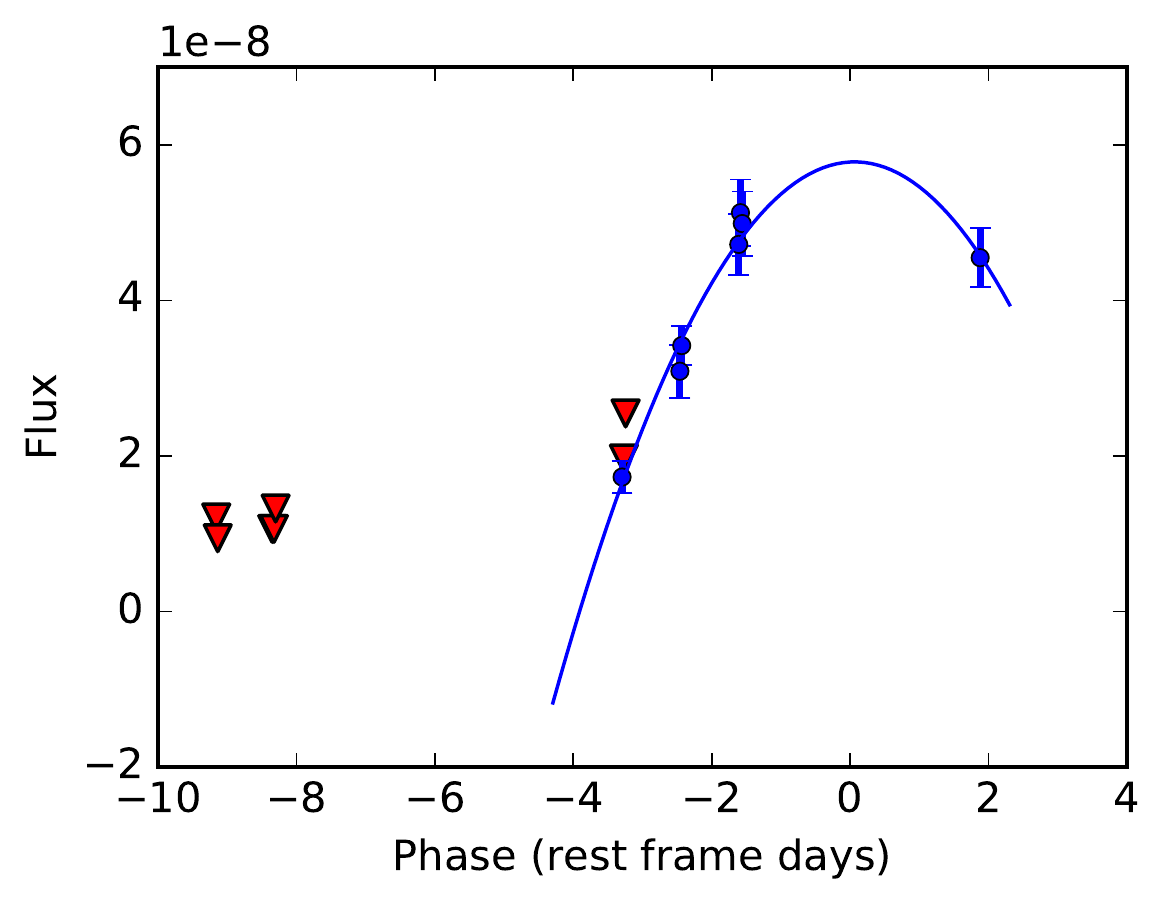}
\caption{Early $g$~band light curve of iPTF\,16asu, showing the rise of the light curve to peak. Red triangles denote $g$~band non-detections. The second-order polynomial best-fit line (blue)} results in a calculated rise time of $3.97 \pm 0.19$~days. The equation of line of best fit is $y=-3.7\times10^{-9}~x^{2}+4.8\times10^{-10}~x+5.8\times10^{-8}$, where $x$ is phase in days and $y$ is flux (F$_{\lambda}$)in arbitrary units.
\label{fig:rise}
\end{figure}

The light curves of iPTF\,16asu are shown in Figure~\ref{fig:16asu_lc}. The rise and peak are only sampled in $g$~band, so we fit a second-order polynomial to the $g$~band light curve near peak brightness to determine a best-fit explosion date, time of peak, and peak luminosity. The fit is shown in Figure~\ref{fig:rise}, and the explosion and best fit peak dates are MJD $57518.53\pm0.17$ and  MJD $57523.25\pm0.14$, respectively. Corresponding calendar dates are 2016 May 10.53 and 2016 May 15.25. Thus, the rise time (time of peak $-$ time of explosion) is $3.97 \pm 0.19$~days in the rest frame. The last optical upper limit prior to the first detection was MJD 57513.31, setting an upper limit to the rise time of 9.94~days. 

Using our series of spectra, we calculate $K$-corrections from the observed filters at $z=0.187$ to rest-frame filters, listed in Table~\ref{tab:kcor}. At this redshift, observed $gri$ corresponds most closely to rest-frame $BVr$, and the wavelength coverage of our spectra allows us to also calculate $K$-corrections to $u$, $g$ and $i$ filters. Applying this, we find that the time of peak corresponds to a peak absolute magnitude of $M_{\rm B} = -20.4$~mag (AB).

\subsection{Light Curve Comparisons}
\label{sec:lc_compare}

iPTF\,16asu inhabits an unusual location in rise time vs. luminosity parameter space (see Figure~\ref{fig:intro}). In this section, we compare its light curve in more detail to objects in the literature that have been noted for their fast timescales and/or high luminosities.
Unfortunately, $K$-corrections are not available for the majority of our comparison objects due to a lack of spectroscopic coverage. For the purposes of comparison, we corrected iPTF\,16asu and all comparison objects for redshift using the following equations:
\begin{equation}
    \lambda = \frac{\lambda_{obs}}{(1+z)}
\end{equation}

\begin{equation}
    M = m_{obs} - 5~\log_{10}{\left(\frac{D_{L}}{10 pc}\right)} + 2.5~\log_{10}({1+z})
\end{equation}
 
\begin{figure*}
\centering
\begin{tabular}{cc}
\includegraphics[width=3.5in]{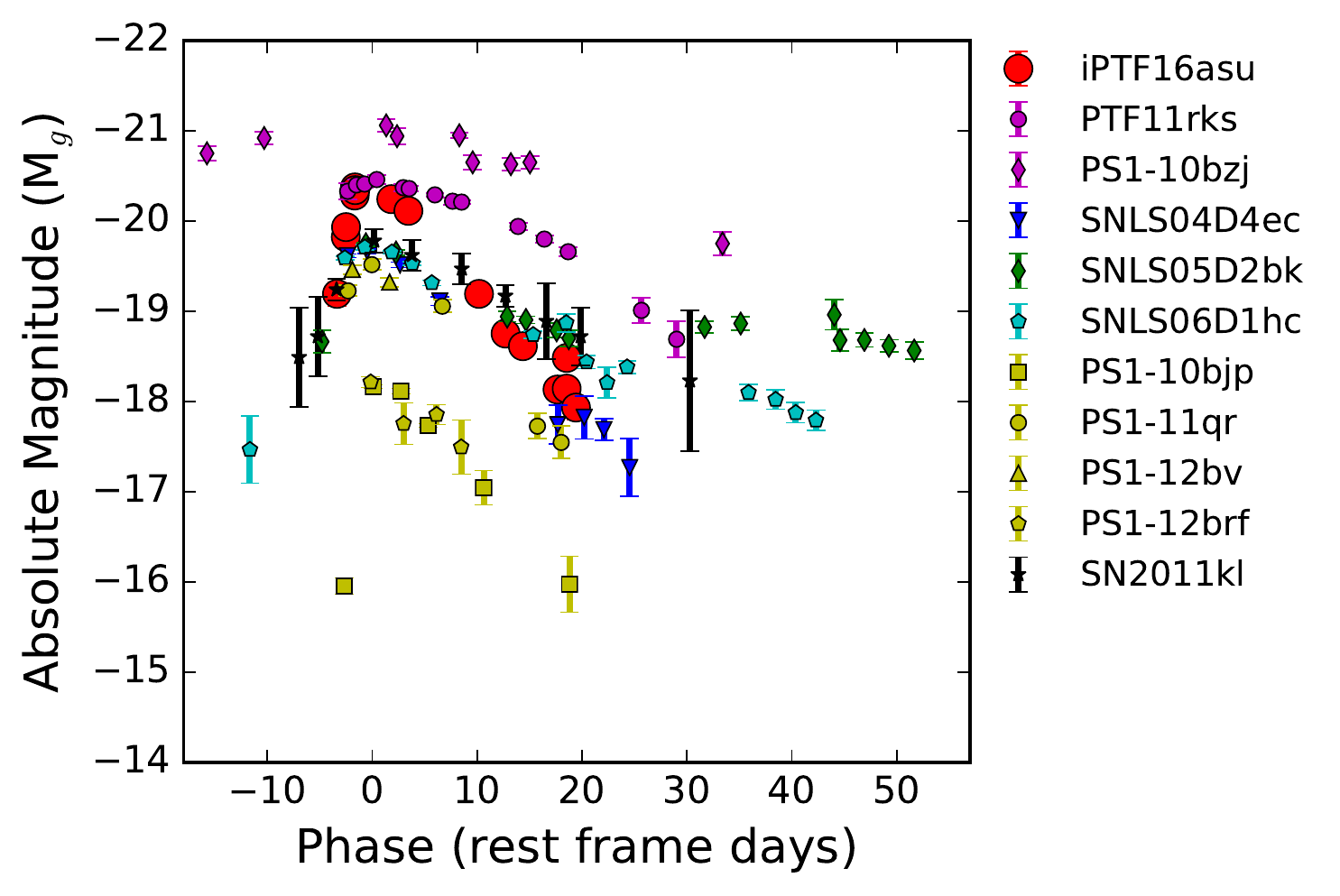} &
\includegraphics[width=3.5in]{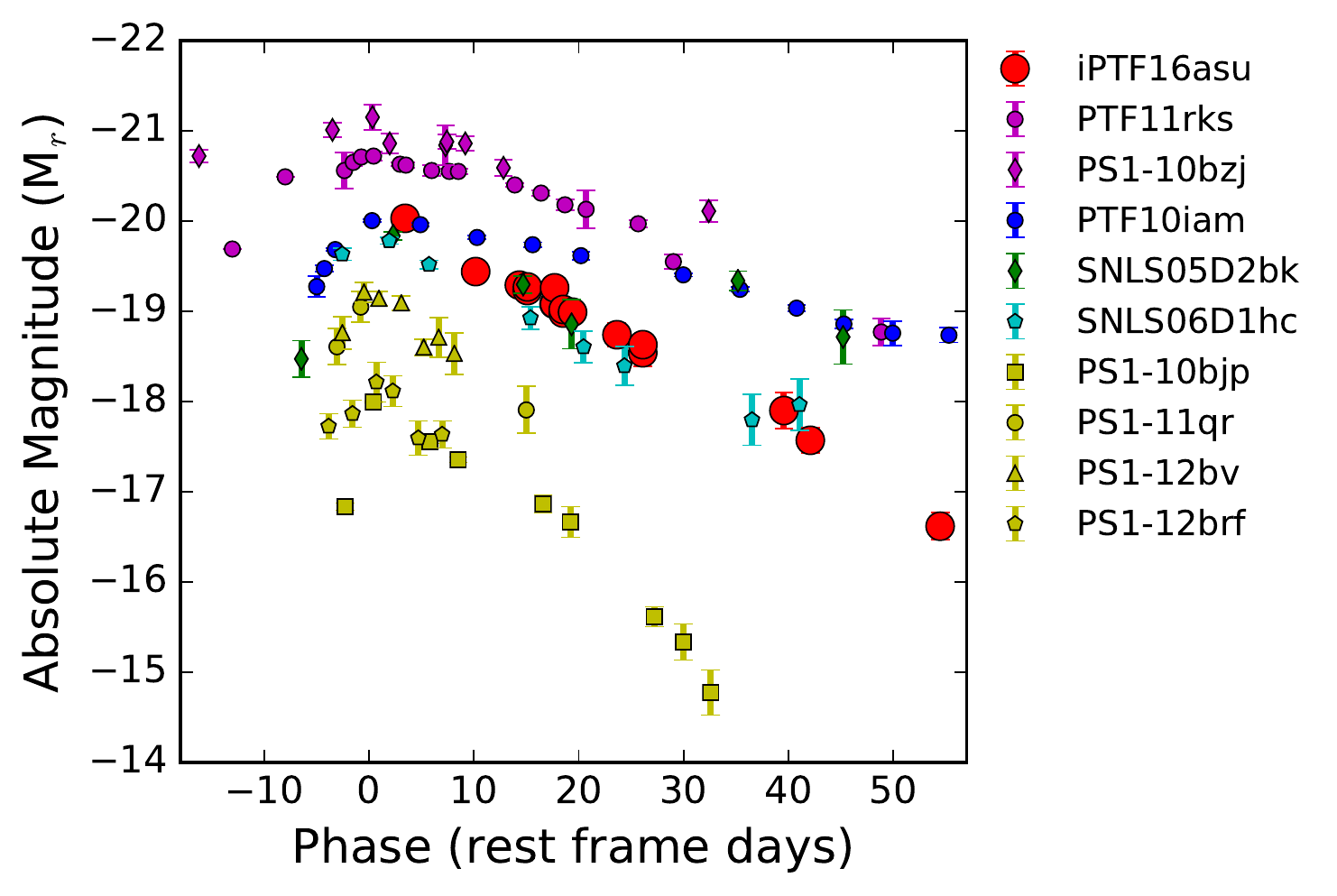}
\end{tabular}
\caption{$g$~band (left) and $r$~band (right) light curve of iPTF\,16asu (red) compared to other luminous and/or rapidly evolving transients from the literature. Filters have been chosen to correspond to approximately the same rest wavelengths. SLSNe from \citet{isj+13} and \citet{lcb+13} shown in magenta; \citet{awh+16} objects shown in blue, cyan, and green; \citet{dcs+14} objects shown in yellow; SN\,2011kl/GRB\,111209A \citep{gmk+15} shown in black.}
\label{fig:lc_comp}
\end{figure*}

\noindent We then choose filters with rest wavelengths as closely corresponding to those of iPTF16asu as possible, in order to facilitate comparison. Comparing this approximation to the actual $K$-corrections calculated for iPTF16asu (Table~\ref{tab:kcor}), we expect this to introduce errors on the order of 0.1~mag. Figure~\ref{fig:lc_comp} shows comparisons to the $g$~band (left) and $r$~band (right) light curves.

First, we compare against the light curves of SNe noted for both their high luminosities and rapid timescales. These include: SN\,2011kl \citep{gmk+15}, a SN associated with the ultra-long gamma-ray burst GRB\,111209A (plotted in black); and PTF\,10iam (blue), SNLS04D4ec (blue), SNLS05D2bk (green) and SNLS06D1hc (cyan) from \citet{awh+16}. In the $g$~band as seen in Figure~\ref{fig:lc_comp} (left), iPTF\,16asu reaches a higher peak luminosity than these transients by over half a magnitude. Measuring from rest-frame phase at $M_{\rm peak -1\rm\,mag}$ to $M_{\rm peak +1\rm\,mag}$, iPTF\,16asu's timescale is about two times shorter with $\tau_{\rm peak -1\rm\,mag}=10$~days. iPTF\,16asu displays both a steeper rise and decay than the \citet{awh+16} objects and SN\,2011kl in the $g$~band. However, iPTF\,16asu resembles these objects more closely in the $r$~band, shown in Figure~\ref{fig:lc_comp} (right). The peak $r$~band magnitude of iPTF\,16asu is approximately the same as PTF\,10iam, SNLS05D2bk, and SNLS06D1hc and the slope of decay runs nearly parallel to that of SNLS06D1hc. Although we have no data on the rise in $r$~band, iPTF\,16asu has a similar peak magnitude to the \citet{awh+16} objects and decays on the same timescale as SNLS06D1hc.

Next we compared the light curve to PS1-10bjp, PS1-11qr, PS1-12bv, and PS1-12brf, a sample of rapidly evolving transients from the Pan-STARRS1 Medium Deep Survey \citep{dcs+14}. The objects shown are the four most luminous objects from the ``gold'' sample, and are plotted in yellow in Figure~\ref{fig:lc_comp}. They have similar rise times and decay slopes to iPTF\,16asu, but are much fainter. In the $g$~band, PS1-11qr and PS1-12bv are the brightest of the Pan-STARRS1 objects reaching a peak magnitude of about $-19.5$~mag; thus iPTF\,16asu is a magnitude brighter at peak. As seen in Figure~\ref{fig:lc_comp}, the shape of iPTF\,16asu's light curve is quite similar to that of PS1-10bjp and PS1-11qr. Early in the decay of iPTF\,16asu, the slope is nearly parallel to that of PS1-10bjp; however, at late times PS1-10bjp decays more sharply than iPTF\,16asu. Comparing these objects to the $r$~band data is less instructive because iPTF\,16asu's rise was not captured in the $r$~band and most of the Pan-STARRS1 objects do not have late-time data.
  
Finally, we compared to the SLSNe PTF11rks \citep{isj+13} and PS1-10bzj \citep{lcb+13}, which are both on the lower-luminosity end of SLSNe. In the $g$~band, iPTF\,16asu reaches about the same peak absolute magnitude as PTF11rks. In the $r$~band iPTF\,16asu's peak luminosity is about $0.5$~mag dimmer than that of PTF11rks. However, the SLSNe have timescales several times longer than iPTF\,16asu, as seen by the much broader peaks. Thus, while iPTF\,16asu reaches similar luminosities as some SLSNe, it evolves on a very different timescale. iPTF\,16asu stands out as a unique and surprising event, even amongst similar transients from the literature.

\subsection{Blackbody Fits}
\begin{figure}
\centering
\includegraphics[width=3.5in]{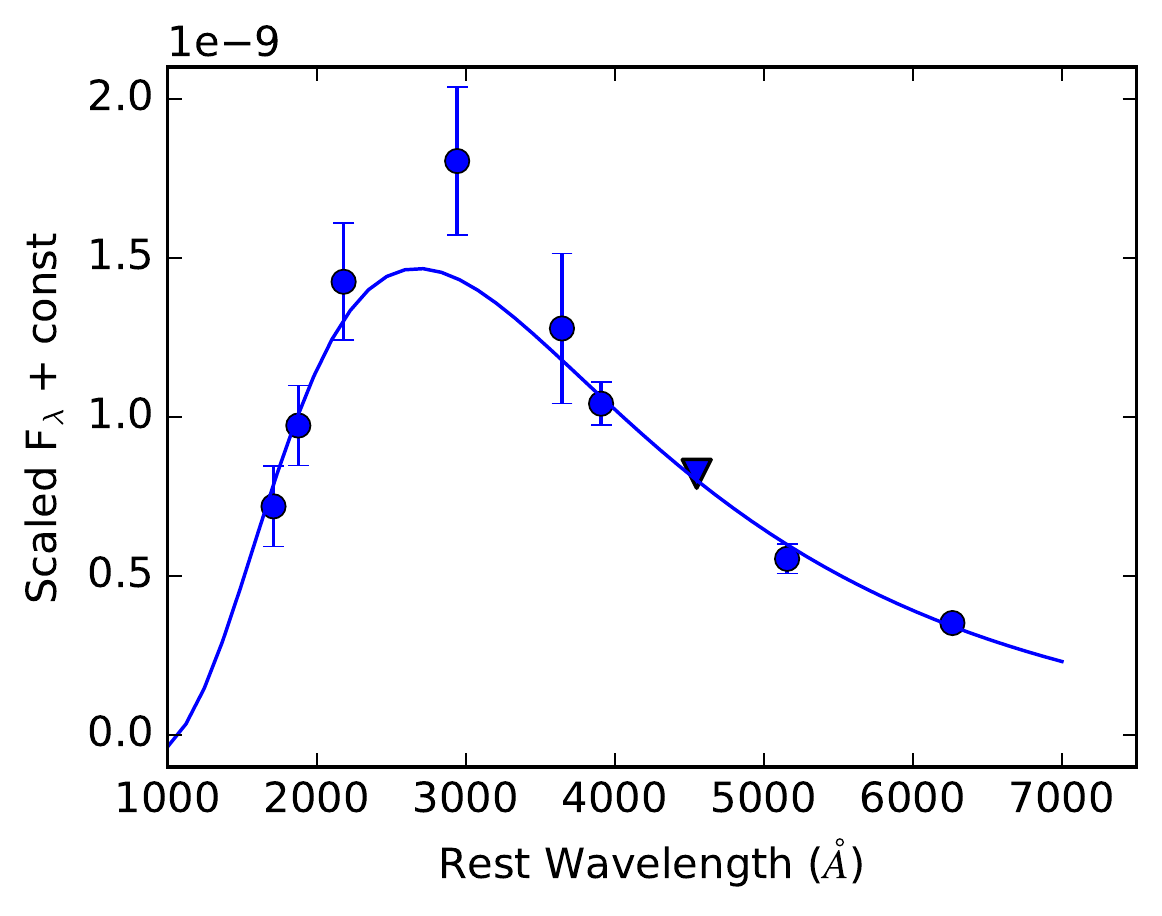}
\caption{Blackbody fit of the \textit{Swift}/UVOT and optical data, at phase $+3$~days. Triangle denotes a non-detection in the V band. The best-fit estimates of the temperature and radius from the fit are T$=10800\pm$250~K and R$=(2.6\pm0.2)\times10^{15}$~cm.
\label{fig:bb_peak}}
\end{figure}

We fit a blackbody to all epochs where we have observations in at least 3 filters, using \texttt{Scipy} least square optimization routines \citep{scipy}, as well as to our two earliest spectra.
Only the day with \textit{Swift}/UVOT detections ($+3$~days past peak) has data in more than 3 filters. The fit to the \textit{Swift} photometry is shown in Figure~\ref{fig:bb_peak}. From this fit we obtain T$=10800\pm250$~K and R$=(2.6\pm0.2) \times10^{15}$~cm. This corresponds to a total blackbody luminosity of $(6.4\pm1.6)\times10^{43}$~ergs~s$^{-1}$.

\begin{figure*}
\centering
\begin{tabular}{cc}
\includegraphics[width=3.5in]{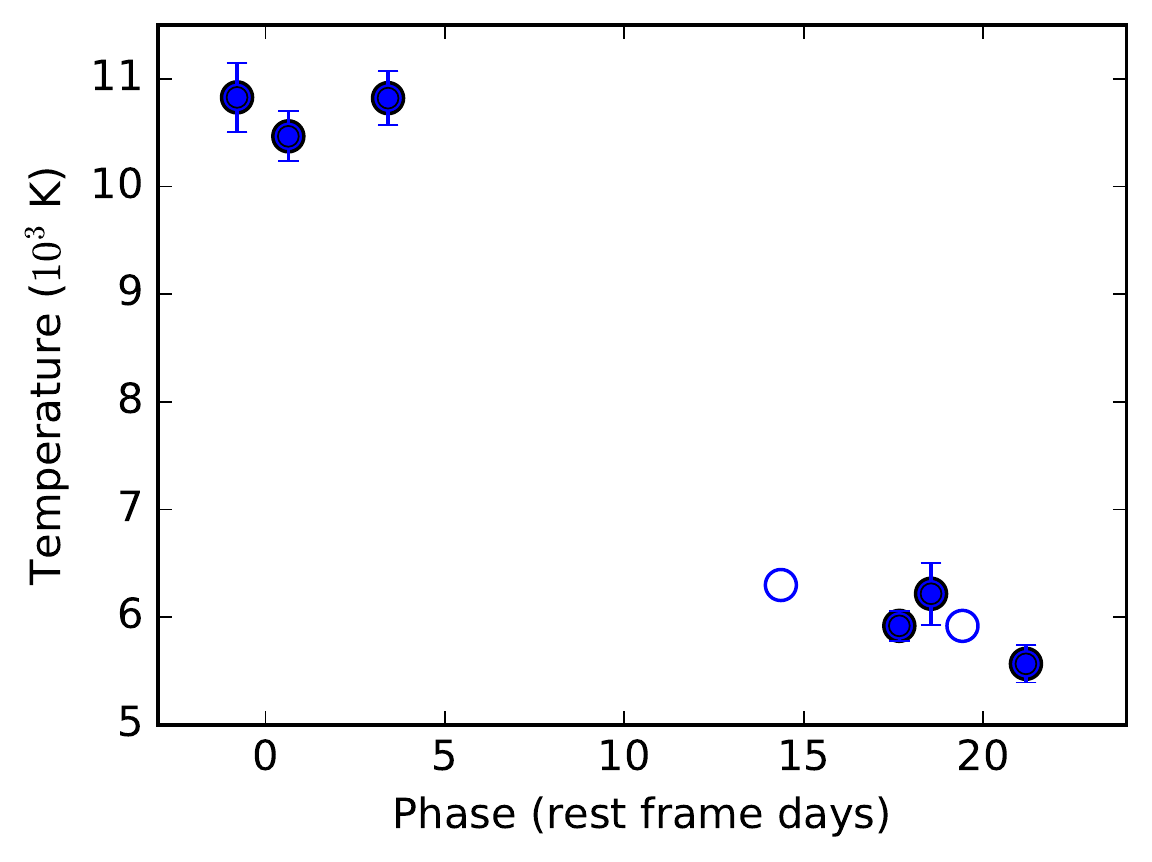} &
\includegraphics[width=3.5in]{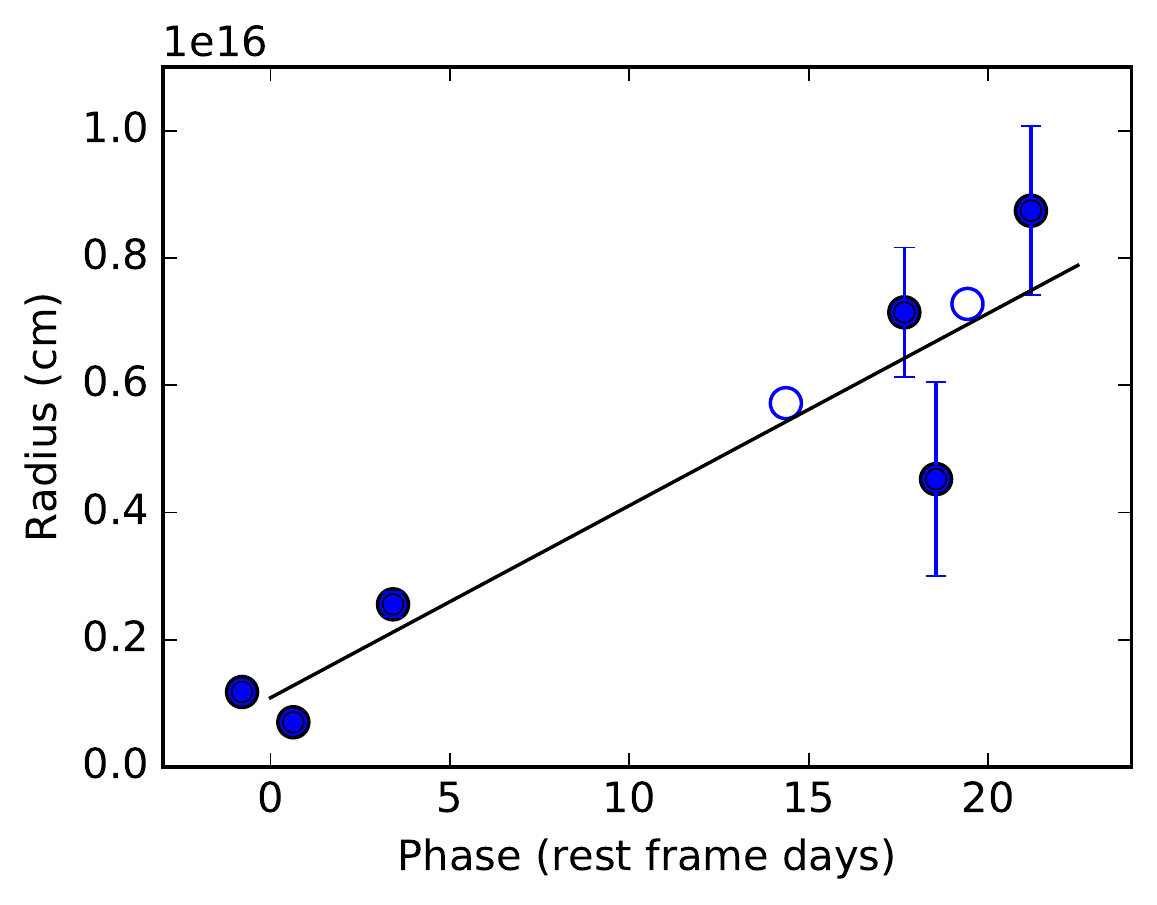}
\end{tabular}
\caption{Blackbody temperature (left) and radius (right) as a function of time. We fit a blackbody to all epochs with photometry in at least three filters, as well as to the earliest two spectra. The slope of the radius over time gives an estimated expansion velocity of $35400\pm5350$~km~s$^{-1}$. Open circles indicate points for which the covariance matrix would not converge to give error bars.}
\label{fig:temprad}
\end{figure*}

Figure \ref{fig:temprad} shows the resulting derived temperatures and radii at all epochs. The overall trends show a cooling blackbody temperature and increasing radius. Fitting a straight line to the blackbody radii, we get a best-fit slope of $34500\pm5400~{\rm km~s}^{-1}$, indicating high average velocities.

\subsection{Bolometric Light Curve}
\label{sec:bol_lc}

We construct a pseudobolometric light curve for iPTF\,16asu by summing the observed flux on days where we have observations in at least three filters. We integrate over the observed spectral energy distribution (SED) using trapezoidal integration, interpolating to the edges of the observed bands. Since this only accounts for the observed flux, it constitutes a strict lower limit on the true bolometric luminosity. 

Pre-peak photometry is only available in the $g$~band so we approximate the rise of the pseudobolometric light curve by assuming a constant ratio of $g$~band flux to total flux, i.e. a constant bolometric correction. This assumption is equivalent to assuming that the temperature on the rise is constant, and equal to the temperature measured from the earliest multiband data. Similarly, for the late-time observations with data only in the $r$~band we estimate the total flux by using the same bolometric correction as from the latest date with data in $\geq3$ filters.

We caution that in constructing a pseudobolometric light curve from optical data only, we are implicitly assuming that the ratio of optical to bolometric flux is approximately constant over the time period of interest. In particular, at late times as the effective temperature falls we would expect the near-IR (NIR) fraction of the bolometric luminosity to increase, which is unconstrained from observations, so assuming a constant bolometric correction is likely an underestimate. Unfortunately, NIR data is also not available for other fast-evolving SNe, so we cannot use them as a basis for comparison. However, given that iPTF\,16asu resembles a normal Type Ic-BL like SN\,1998bw from $\sim 20$ days past explosion onwards (Sections~\ref{sec:spec_analysis}, \ref{sec:nickel}), we expect its late-time evolution of the optical-to-bolometric flux ratio to resemble normal Type Ic-BL SNe. Comparing to the light curve samples analyzed in \citet{lbj14}, we estimate that this adds an uncertainty on the level of 10-20\% at late times.

\begin{figure*}
\centering
\begin{tabular}{cc}
\includegraphics[width=3.5in]{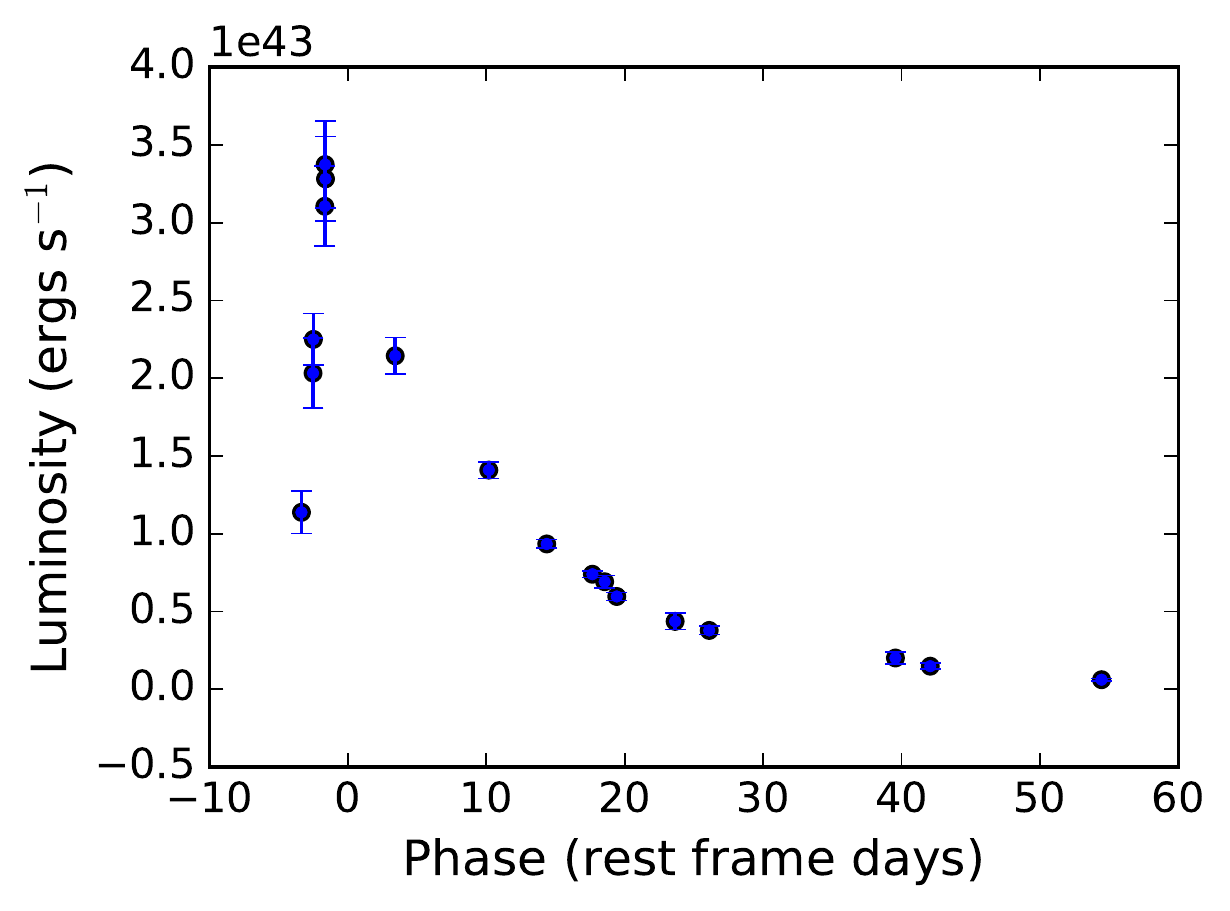} &
\includegraphics[width=3.5in]{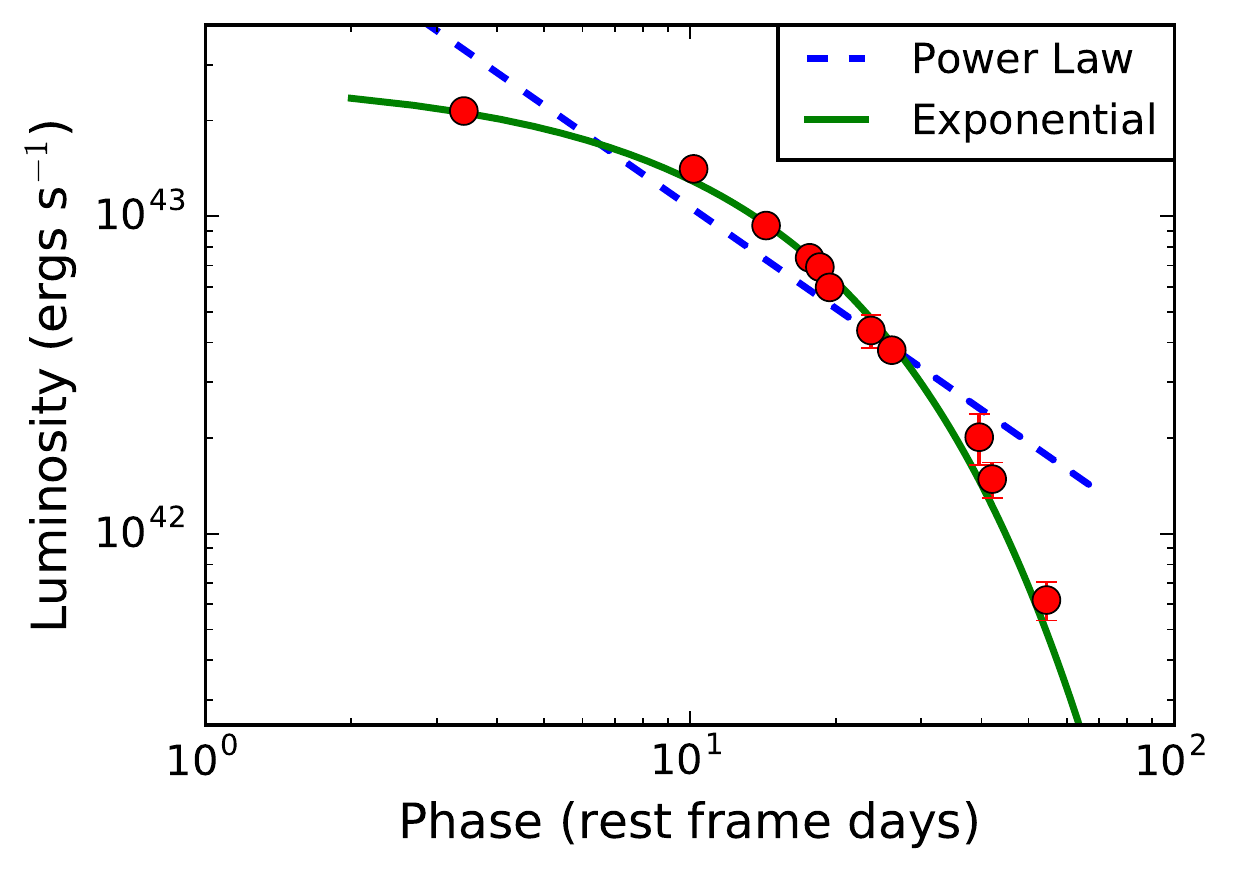}
\end{tabular}
\caption{\textit{Left:} pseudobolometric light curve of iPTF\,16asu. Luminosities obtained from data using the trapezoidal integration method. \textit{Right:} Fit of the decline of iPTF\,16asu's light curve to a power law and an exponential. The power law (dashed blue) has a best-fit of $L~\propto~t^{-1.06 \pm 0.14}$ and the exponential (solid green) decays on a timescale of $\tau=13.56\pm0.56$~days. The light curve decline is well fit by an exponential.}
\label{fig:lum}
\end{figure*}

Figure~\ref{fig:lum} (left) shows the resulting pseudobolometric light curve. Using trapezoidal integration over time we calculate an estimated total radiated energy of  $(4.0\pm0.6)\times10^{49}$~ergs and a peak luminosity of $(3.4\pm0.3)\times10^{43}$~ergs~s$^{-1}$.

The shape of the decline of the bolometric light curve sheds light on what physical processes may be powering this event. Figure~\ref{fig:lum} (right) shows the best-fit power law and exponential fits to the post-peak light curve. Clearly, the decline of the light curve does not follow a power law; however, it fits an exponential well. The power law has a best-fit decay of $L~\propto~t^{-1.06\pm0.14}$ and the exponential decays on a timescale of $\tau=13.56\pm0.56$~days. The power law decay parameter is similar to those found for the objects in \citet{awh+16}. Also similar, two of the four \citet{awh+16} objects are better fit by an exponential. The implications of these results are discussed in Section~\ref{sec:models}.

\section{Spectroscopic Properties}
\label{sec:spec_analysis}

\subsection{Spectroscopic Evolution \& Comparisons}

We obtained eight spectra of iPTF\,16asu between May 14, 2016 and July 6, 2016. The spectra are shown in Figure~\ref{fig:spectra}. In this section, we look at the spectroscopic evolution in more detail, and compare the spectroscopic properties of iPTF\,16asu to similar objects from the literature.

\begin{figure}
\centering
\includegraphics[width=3.5in]{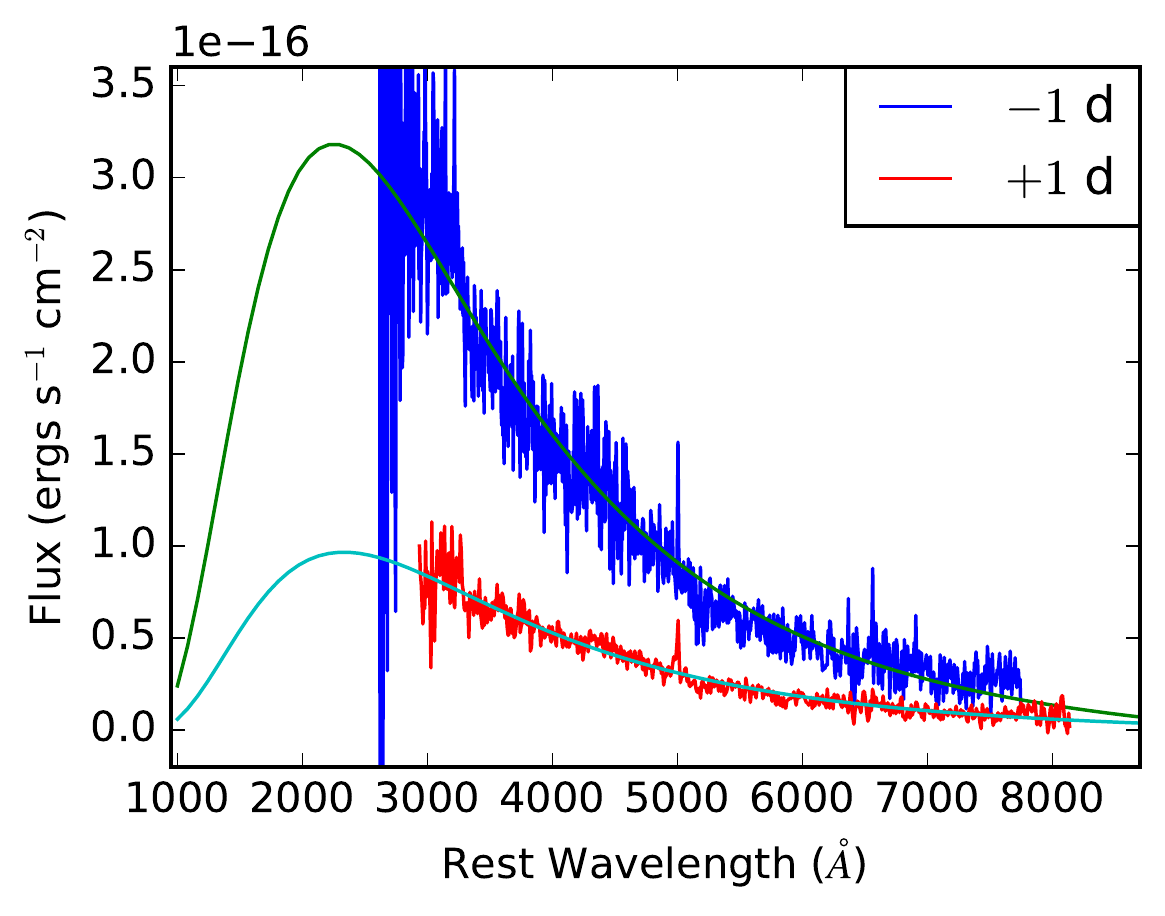}
\caption{Blackbody fit of the two earliest spectra. The corresponding temperature and radius at phase $-1$~day are T=10828$\pm$322 K and R$=(1.18\pm0.07)\times 10^{15}$~cm. The corresponding temperature and radius at phase $+1$~day are T$=10466\pm232$~K and R$=(7.05\pm0.03)\times10^{14}$~cm.
\label{fig:bb_spectra}}
\end{figure}

\begin{figure}
\centering
\includegraphics[width=3.5in]{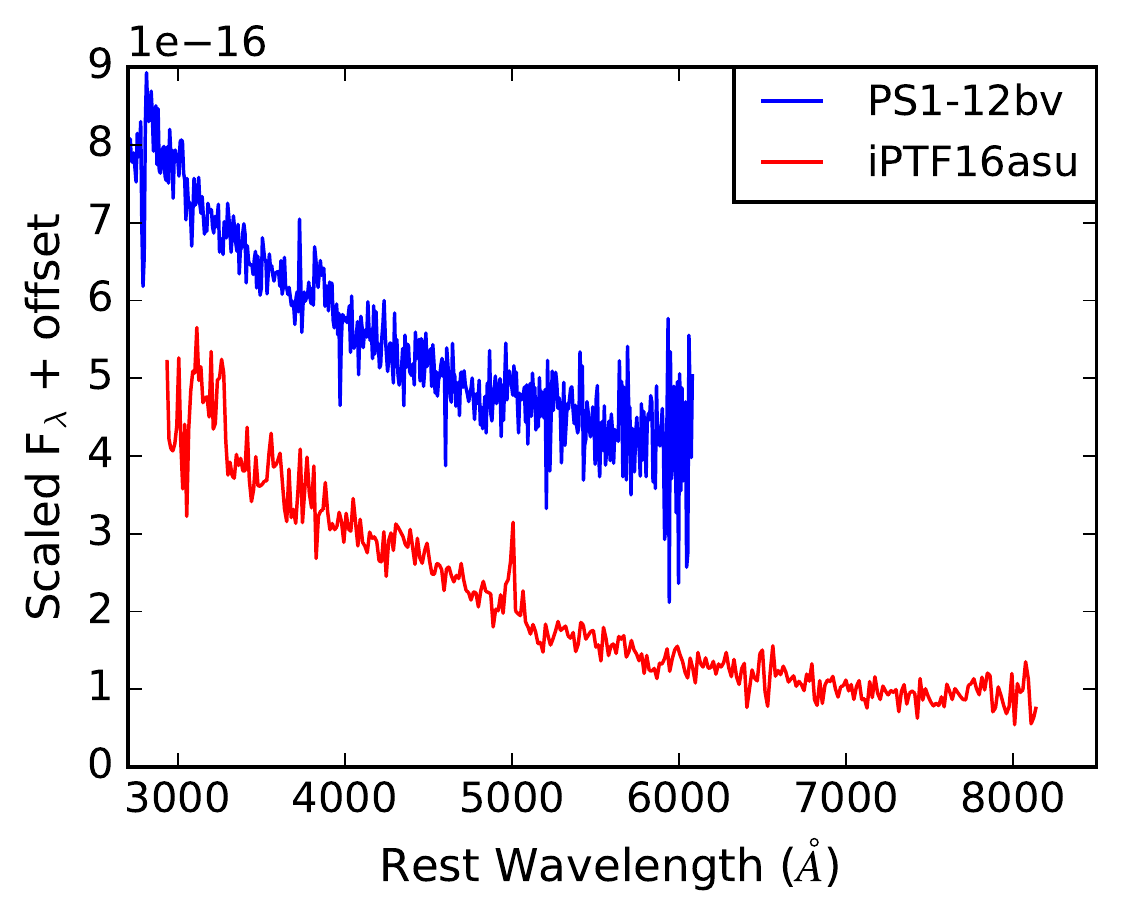}
\caption{Spectrum of PS1-12bv \citep{dcs+14} at +7 d after explosion compared to iPTF\,16asu at +5 d after explosion. iPTF\,16asu spectrum from NOT. Host galaxy narrow emission lines have not been removed -- note the feature at $\sim 5000$~\AA\, in the iPTF\,16asu spectrum is narrow [\ion{O}{3}] $\lambda\lambda$ 4959,5007 emission from the host galaxy that appears broadened here due to binning. 
\label{fig:earlyspec}}
\end{figure}

The first two spectra, taken within a day before and after peak, show a featureless blue continuum with no discernible broad features. The spectrum is well-fit by a blackbody, as shown in Figure~\ref{fig:bb_spectra}. Such spectra dominated by blue continua have also been observed at early phases in other supernovae, typically while the luminosity is powered by cooling of the stellar envelope following shock breakout (see e.g. SN\,1993J; \citealt{wew+94,mfh+00}).
Interestingly, the rapidly evolving SNe from Pan-STARRS1 \citep{dcs+14}  also showed featureless, blue continua. 
Figure~\ref{fig:earlyspec} shows a comparison of PS1-12bv at peak compared to iPTF\,16asu at peak. Unfortunately, comparison at late times is not possible, as there is no further follow-up spectroscopy on the Pan-STARRS events. Based on the limited spectroscopic data available we cannot rule out that they were caused by the same phenomenon as iPTF\,16asu.

\begin{figure}
\centering
\includegraphics[width=3.5in]{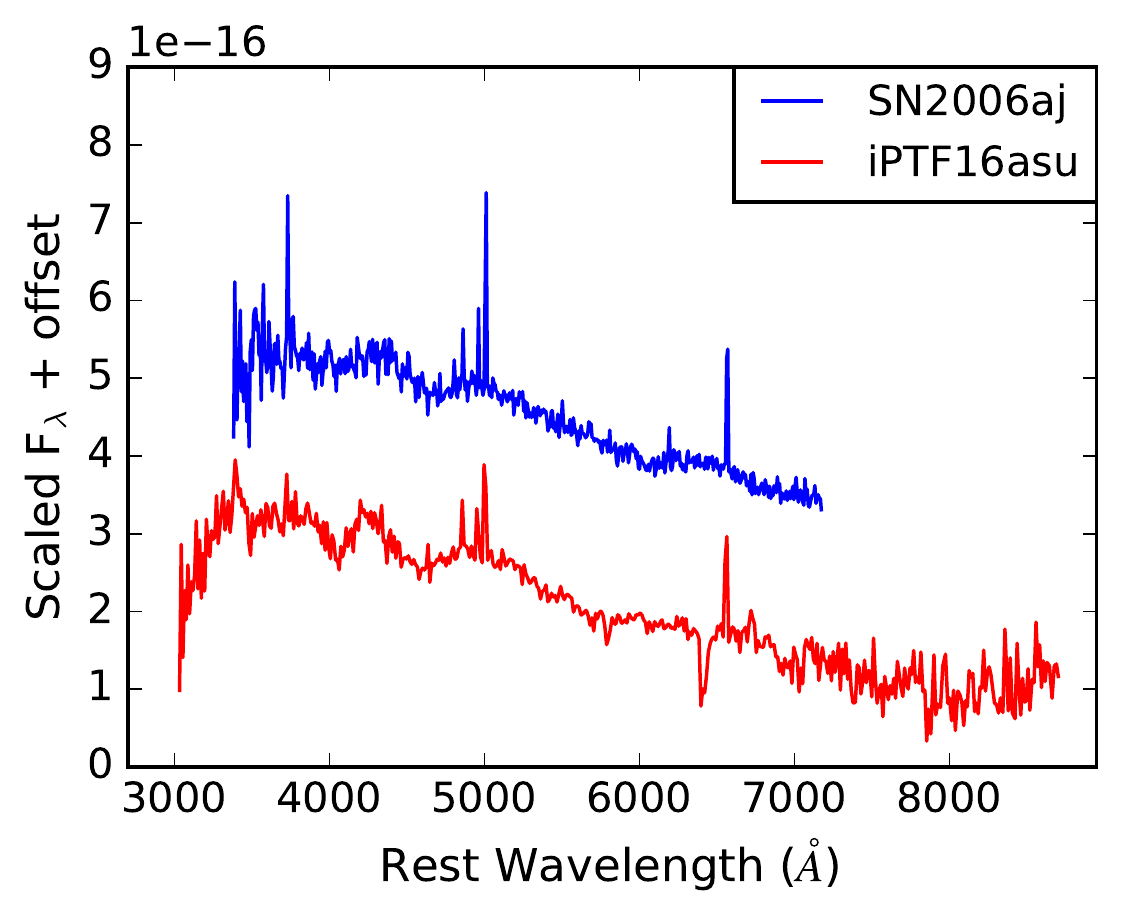}
\caption{Spectrum of SN\,2006aj \citep{msg+06} at +6 d after explosion compared to iPTF\,16asu at +12 d after explosion. iPTF\,16asu spectrum from TNG. Host galaxy narrow emission lines have not been removed.
\label{fig:midspec}}
\end{figure}

The next two spectra, taken at phases 8 and 10 days past maximum, still show an underlying blue continuum, but with broad features emerging. Such an evolution is reminiscent of GRB-SNe. To illustrate this we show a comparison to SN\,2006aj/GRB\,060218 \citep{msg+06} in Figure~\ref{fig:midspec}. SN\,2006aj is of particular interest here because it is one of the few GRB-SNe that would not be ruled out by our radio and X-ray limits. We discuss GRB models for iPTF\,16asu in detail in Section~\ref{sec:grb}.

\begin{figure}
\centering
\includegraphics[width=3.5in]{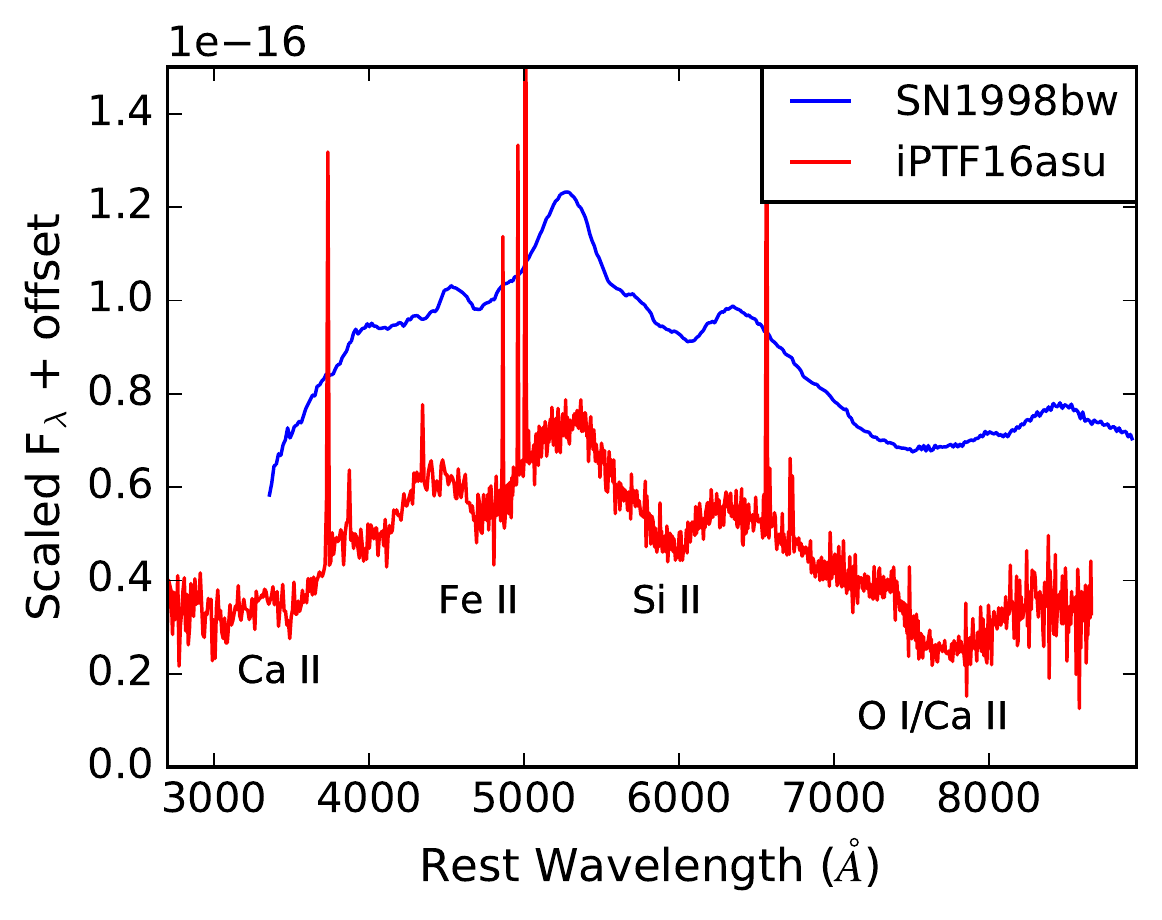}
\caption{Spectrum of SN\,1998bw \citep{pcd+01} at +18 d after explosion compared with iPTF\,16asu at +23 d after explosion. Features commonly identified in SNe Ic-BL are marked. iPTF\,16asu spectrum from  Keck1+LRIS. Host galaxy narrow emission lines have not been removed.
\label{fig:latespec}}
\end{figure}

The three spectra taken at phases $+17$, $+19$ and $+22$~days post maximum are dominated by distinct, broad-line features, leading us to classify iPTF\,16asu as a Type Ic-BL SN. Figure~\ref{fig:latespec} shows a comparison of iPTF\,16asu at $+23$~days after explosion ($+19$~days past peak) to SN\,1998bw at $+18$~days after explosion, and features commonly identified in Type Ic-BL SNe are marked. Interestingly, the spectra of these events look very similar at roughly the same time after explosion, suggesting that iPTF\,16asu may have a  normal-timescale supernova component hidden underneath the luminous and rapidly-evolving peak.

It is also worth noting that the spectroscopic evolution of iPTF\,16asu is different from the few objects in \citet{dcs+14} and \citet{awh+16} with spectra at later phases: PS1-12bb showed a featureless continuum at phase $+33$~days, while PTF\,10iam showed broad H$\alpha$ emission at phase $+28$~days. This spectroscopic diversity suggests that there are likely multiple physical mechanisms giving rise to light curves in this part of transient phase space.

Our final spectrum, taken at a phase $+44$~days past peak, is dominated by host galaxy light. We discuss the host galaxy properties in Section~\ref{sec:hostgal}. 

\subsection{Velocities}

Measuring velocities from Type Ic-BL spectra is challenging, since the lines are often blended due to the high velocities. In addition, different lines can give different velocities because these elements are formed and found at different radii in the expanding, ejected material. For iPTF16asu, we choose the strongest lines which are the \ion{Si}{2}~$\lambda6355$\,\AA~ line and the \ion{Fe}{2}~$\lambda5169$\,\AA~ line.

In the case of the \ion{Si}{2}~$\lambda6355$\,\AA~ line we fit a parabola to find the minimum of the broad absorption feature. The corresponding wavelength is then used to determine velocities using the relativistic Doppler shift. The measured velocities are listed in Table~\ref{tab:vel}.

\begin{figure}
\centering
\includegraphics[width=3.5in]{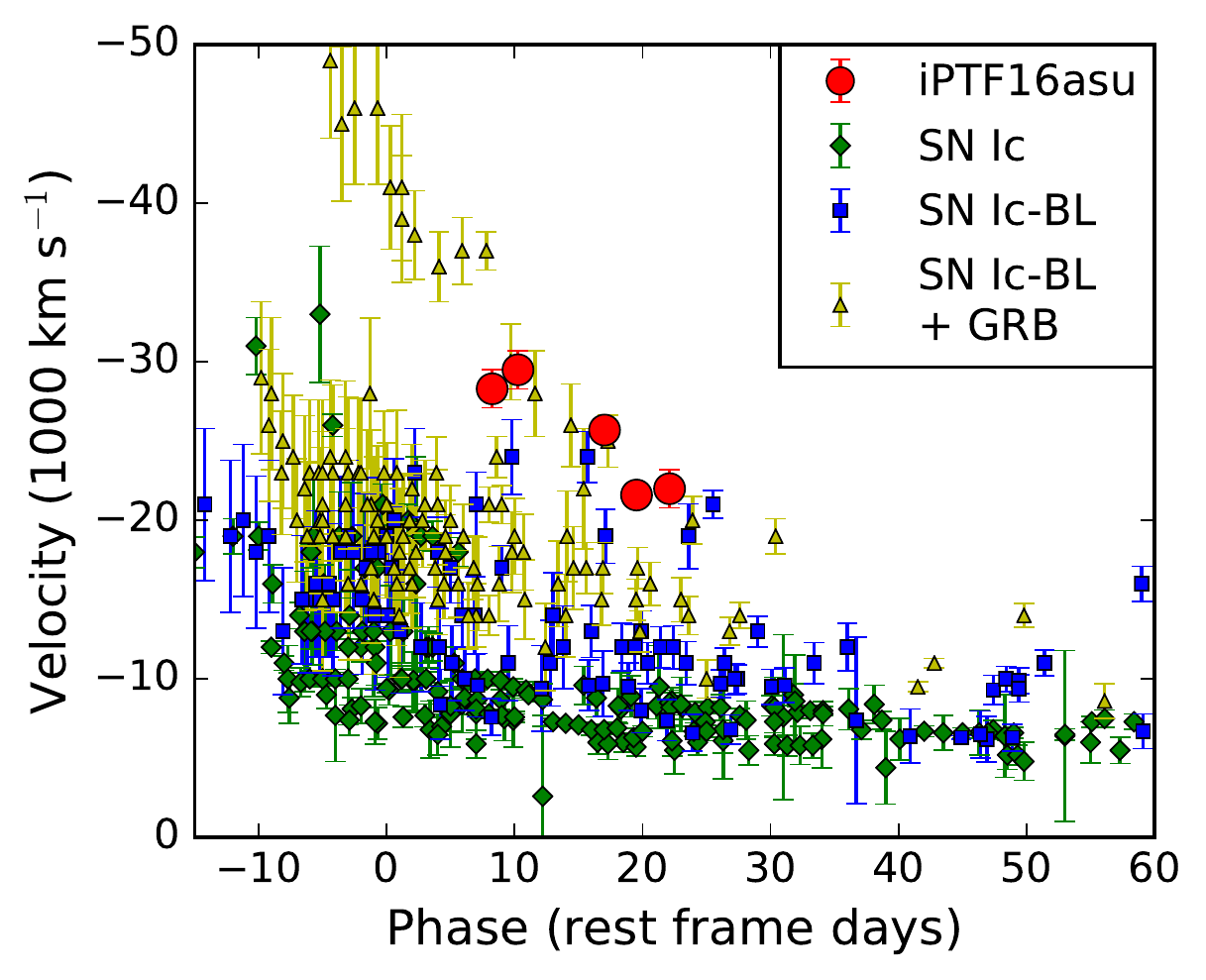}
\caption{Velocity evolution of iPTF\,16asu, measured from the  \ion{Fe}{2}~$\lambda5169$\,\AA~ line (red points), compared against literature data of SNe Ic (green diamonds), SNe Ic-BL (blue squares), and SNe Ic-BL (yellow triangles) associated with GRBs. Data from \citet{mlb16}.
\label{fig:FeII_vels}}
\end{figure}

In the case of the \ion{Fe}{2}~$\lambda5169$\,\AA~ line, similar to other Type Ic-BL SNe, this line is blended with the neighboring \ion{Fe}{2}~$\lambda$4924\, and \ion{Fe}{2}~$\lambda$5018\, lines. Thus, we cannot simply fit the minimum of this feature to derive velocities. Instead, we use the convolution method developed by \citet{mlb16} and \citet{lmb+16} to extract velocities from the \ion{Fe}{2}~$\lambda5169$\,\AA~ line. The measured velocities are listed in Table~\ref{tab:vel}. Figure~\ref{fig:FeII_vels} shows the \ion{Fe}{2} velocities from iPTF\,16asu compared to the sample of Ic and Ic-BL SNe from \citet{mlb16}, with velocities derived using the same method (and code). The \ion{Fe}{2} velocities we measure for iPTF\,16asu are high compared to the objects in this sample, closest to the \ion{Fe}{2} velocities of SNe Ic-BL associated with GRBs. We note that phase in this figure is measured with respect to maximum light -- if iPTF\,16asu has a ``normal'' SN component hidden underneath the blue, luminous peak, the supernova maximum would be later and iPTF\,16asu would move left in this plot, but the basic conclusion that the velocities are comparable to SN Ic-BL with associated GRBs would be unchanged.

\section{Host Galaxy}
\label{sec:hostgal}

The host galaxy of iPTF\,16asu is detected both in the PTF templates and in SDSS images. The observed SDSS model magnitudes are $u' = 22.90 \pm 0.37$~mag, $g' = 22.10 \pm 0.09$~mag, $r' = 21.82 \pm 0.11$~mag, $i' = 21.43 \pm 0.11$~mag, and $z' = 21.25 \pm 0.28$~mag. At a redshift of $z=0.1874$, this makes the host a dwarf galaxy, with an absolute magnitude $M_g \simeq -17.5~{\rm mag}$. We use the {\tt FAST} code \citep{kvl+09} to fit a galaxy model to the observed photometry, using a \citet{mar05} stellar population synthesis model, and assuming a Salpeter IMF and an exponential star formation history. The metallicity and extinction are constrained by our spectroscopic data (see below), so we use the extinction derived from the Balmer decrement, and a metallicity of $Z=0.5~Z_{\odot}$, which is the closest model grid value to our derived metallicity. With these assumptions, we find a best-fit stellar mass of $M_{*}=4.6_{-2.3}^{+2.0} \times 10^8~{\rm M}_{\odot}$ and a best-fit stellar population age $5.0_{-4.6}^{+6.5} \times 10^8~{\rm yr}$.

\begin{figure}
    \centering
    \includegraphics{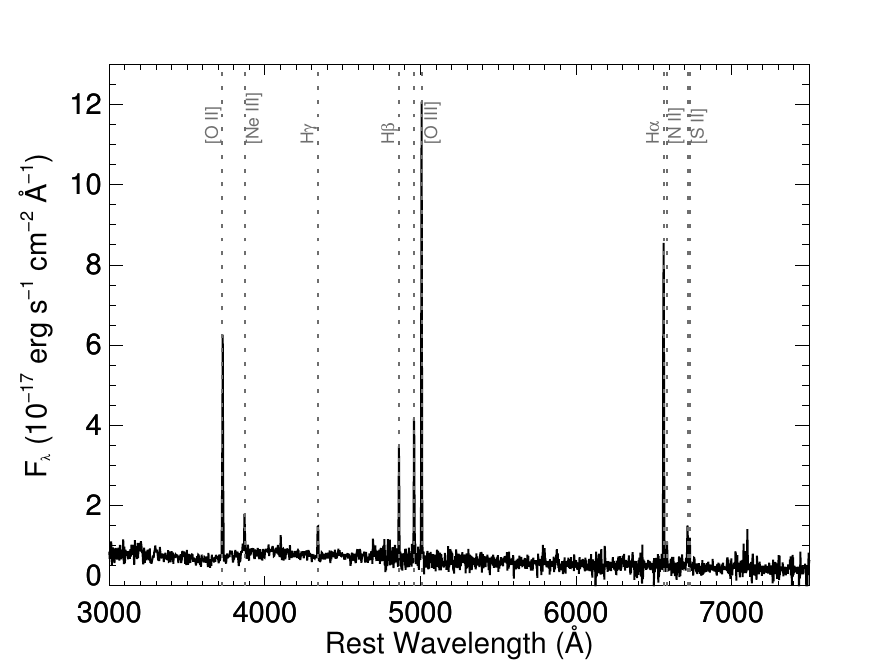}
    \caption{Spectrum of the host galaxy of iPTF\,16asu, taken with Keck1+LRIS at $\sim 300$~days after the SN explosion. Strong galaxy emission lines are marked.}
    \label{fig:hostspec}
\end{figure}

We obtained a host galaxy spectrum nearly a year after explosion, shown in Figure~\ref{fig:hostspec}. We scale this galaxy spectrum to the SDSS photometry to account for slit losses, and measure the fluxes of the (unresolved) lines by fitting Gaussian profiles. The measured emission line fluxes are listed in Table~\ref{tab:lineflux}.

We use the Balmer decrement to calculate the host galaxy extinction, assuming Case B recombination \citep{ost89}. We measure a ${\rm H}\alpha / {\rm H}\beta$ ratio of $3.5 \pm 0.2$, translating to a host extinction ${\rm E}(B-V)= 0.22 \pm 0.06$, assuming a standard Milky Way extinction curve with $R_V = 3.1$ \citep{ccm89}. Using the extinction-corrected $H\alpha$ flux, we measure a star formation rate of 0.7~M$_{\odot}~{\rm yr}^{-1}$ \citep{ken98}. Given the stellar mass derived from the photometry, this corresponds to a specific star formation rate of $1.4~{\rm Gyr}^{-1}$.

We use {\tt pyMCZ} \citep{bmo+16} to calculate the galaxy oxygen metallicity from the [\ion{O}{3}], [\ion{O}{2}], [\ion{N}{2}], H$\alpha$ and H$\beta$ lines. {\tt pyMCZ} is a Python-based implementation of up to 15 metallicity calibrators, updating the code given in \citet{kd02} and \citet{ke08} and with better treatment of statistical uncertainty from Monte Carlo sampling. While there is some scatter between the different strong-line metallicity estimators, they generally agree that the host galaxy of iPTF\,16asu is low metallicity. For example, we find values of $12+\log({\rm O/H})$ to be $8.12_{-0.07}^{+0.04}$ on the \citet{pp04} O3N2 scale, $8.22_{-0.07}^{+0.18}$ on the \citet{mcg91} scale, and $8.39_{-0.05}^{+0.11}$ on the \citet{kk04} R$_{23}$ scale, to name three commonly used indicators. Using a solar oxygen abundance of $12+\log({\rm O/H}) = 8.69 \pm 0.05$ \citep{ags+09}, this translates to a metallicity $Z \simeq Z_{\odot}/3$. 

Taken together, the host galaxy of iPTF\,16asu was a low-mass, low-metallicity, starforming dwarf galaxy. Such an environment is not unusual for SNe Ic-BL, which, in general, are found in lower-metallicity galaxies than other stripped-envelope SNe; for example, the median metallicity of SN Ic-BL hosts in the compilation of \citet{ssl+12} was $12+\log({\rm O/H}) = 8.20$ on the \citet{pp04} O3N2 scale. Other rare transients, such as long GRBs and SLSNe also show a preference for low-metallicity galaxies \citep[e.g.,][]{lbk+10,lcb+14,pqy+16}. The high specific star formation rate and the strong [\ion{O}{3}]$\lambda$5007\,\AA\, line (${\rm EW}_{5007} \simeq 87~{\rm \AA}$, rest-frame), in particular, is reminiscent of SLSN host galaxies \citep{lsk+15}. Interestingly, the same is not true for the rapidly evolving transients studied in \citet{dcs+14} and \citet{awh+16}: for both samples, the host galaxies were generally more massive galaxies near solar metallicity.

\section{Model Comparisons}
\label{sec:models}

\subsection{Nickel Decay}
\label{sec:nickel}
Most SNe Ic/Ic-BL are powered by the release of energetic photons from the radioactive decay of $^{56}$Ni into $^{56}$Co and finally $^{56}$Fe. Since the late time spectra of iPTF\,16asu look very similar to the spectra of other SNe Ic-BL (Section~\ref{sec:spec_analysis}), we first consider whether the light curve of iPTF\,16asu can be explained purely by the decay of $^{56}$Ni.

\begin{figure}
\centering
\includegraphics[width=3.5in]{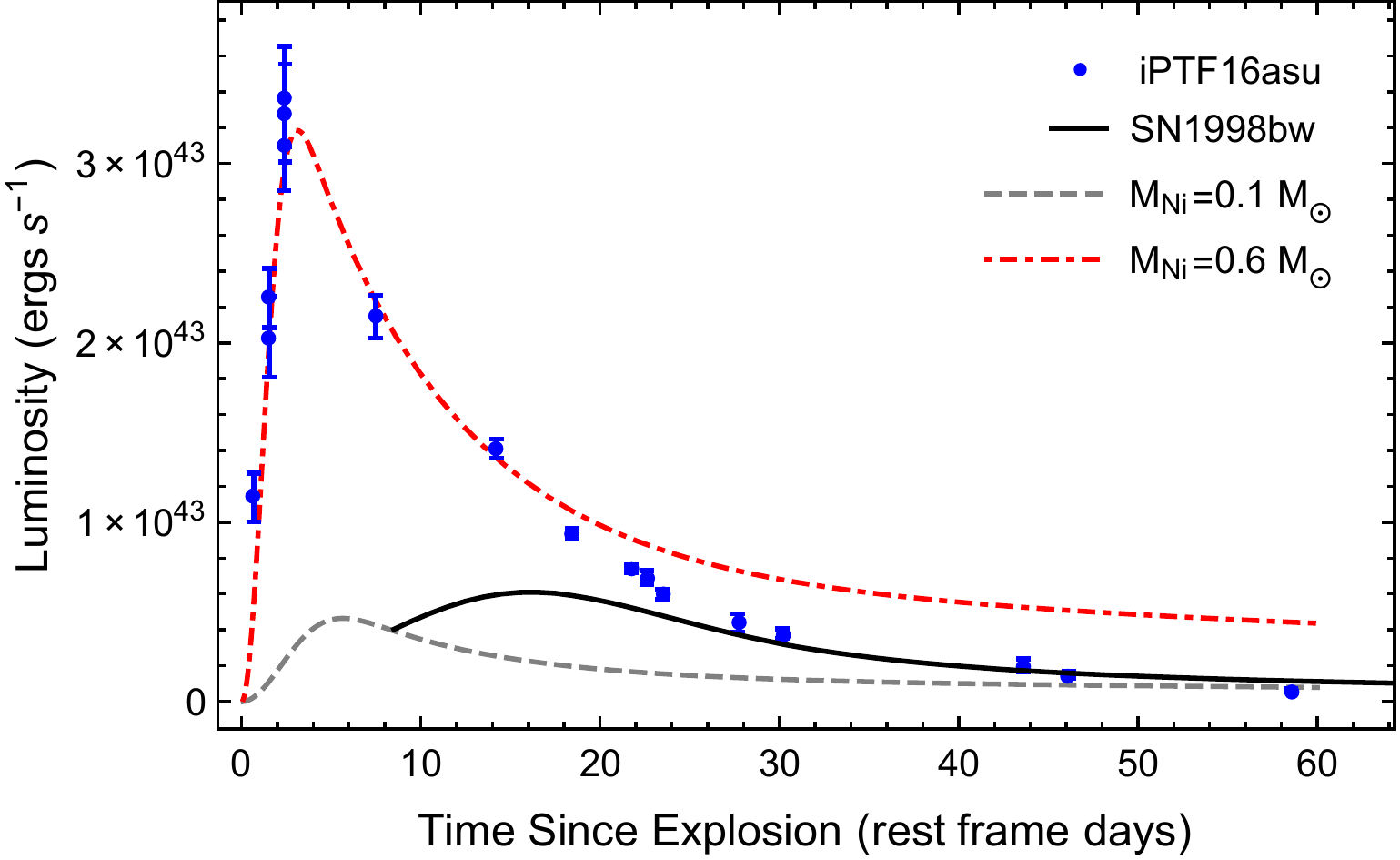}
\caption{Nickel-powered model fit to the light curve of iPTF\,16asu, following \citet{arn82} and \citet{lbj+16}. The red dot-dashed curve shows the model with parameters $M_{\rm Ni}=0.55~$M$_{\odot}$ and $\tau_{\rm diff}=1.5$~days. The dotted gray line shows the model constrained by the last point with parameters $M_{\rm Ni}=0.1~$M$_{\odot}$ and $\tau_{\rm diff}=3.7$~days. For comparison, the bolometric light curve using $BV(RI)_{\rm c}$~bands of SN\,1998bw \citep{csc+11} is plotted in black. Attempting to fit the sharp, luminous light curve with a $^{56}$Ni model leads to an unphysical solution in which the derived ejecta mass is lower than the required nickel mass.
\label{fig:nickel}}
\end{figure}

Using the equations from Section~2 of \citet{lbj+16}, we compare our pseudobolometric light curve from Section~\ref{sec:bol_lc} to the theoretical model for a $^{56}$Ni decay powered light curve in \citet{arn82}. The model takes two input parameters, diffusion time and $^{56}$Ni mass. The $^{56}$Ni mass predominantly affects the luminosity of the light curve, as a larger $^{56}$Ni mass would indicate a larger total energy input, and the diffusion time controls the timescale over which the energy diffuses out, or the width of the peak. Figure~\ref{fig:nickel} shows the bolometric light curve of iPTF\,16asu plotted against an \citet{arn82} model with parameters $M_{\rm Ni}=0.55~$M$_{\odot}$ and $\tau_{\rm diff}=1.5$~days, assuming an opacity of $\kappa=0.1$~cm$^{2}$~g$^{-1}$. As seen in Figure~\ref{fig:nickel} , the $^{56}$Ni decay model does not fit both the sharp peak and steep decay well, though we caution that the lack of NIR data could mean our late-time bolometric light curve is systematically underestimated (Section~\ref{sec:bol_lc}).

An ejecta mass of M$_{\rm ej}=0.06~$M$_{\odot}$ was calculated using this diffusion time along with an estimate of the kinetic energy. Since our spectra near peak are featureless, and thus we cannot measure a velocity, we used the average velocity ($35000~{\rm km~s}^{-1}$) derived from the evolution of the blackbody radii to calculate this kinetic energy. The ejecta mass is notably about ten times smaller than the amount of $^{56}$Ni required to power this light curve, which is unphysical: the $^{56}$Ni mass cannot be larger than the total ejecta mass, since it is necessarily part of the ejecta. Thus, we rule out spherically symmetric radioactive $^{56}$Ni decay as the dominant energy source for iPTF\,16asu.
 
The Arnett model considered above assumes spherical symmetry and a central energy source, i.e. that all the nickel is in the center. Therefore, we cannot rule out the possibility of $^{56}$Ni-powered models for iPTF\,16asu in a highly mixed or strongly asymmetric scenario (e.g., a jet), though we note that assymmetry is not expected to have a large effect on the observed luminosity \citep{bdl+17}. More sophisticated modeling is outside of the scope of this paper.

\begin{figure*}
\centering
\begin{tabular}{cc}
\includegraphics[width=3.5in]{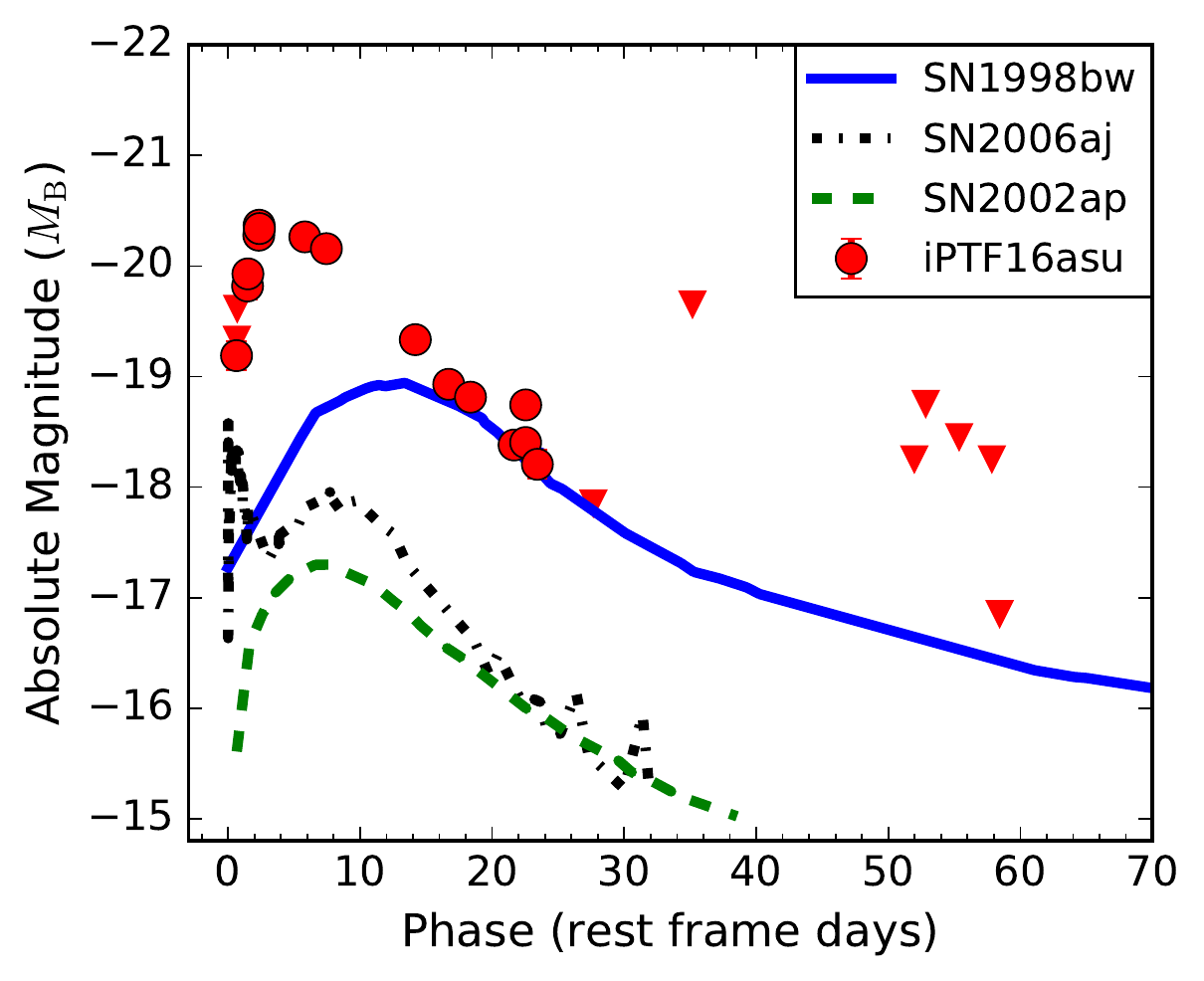} &
\includegraphics[width=3.5in]{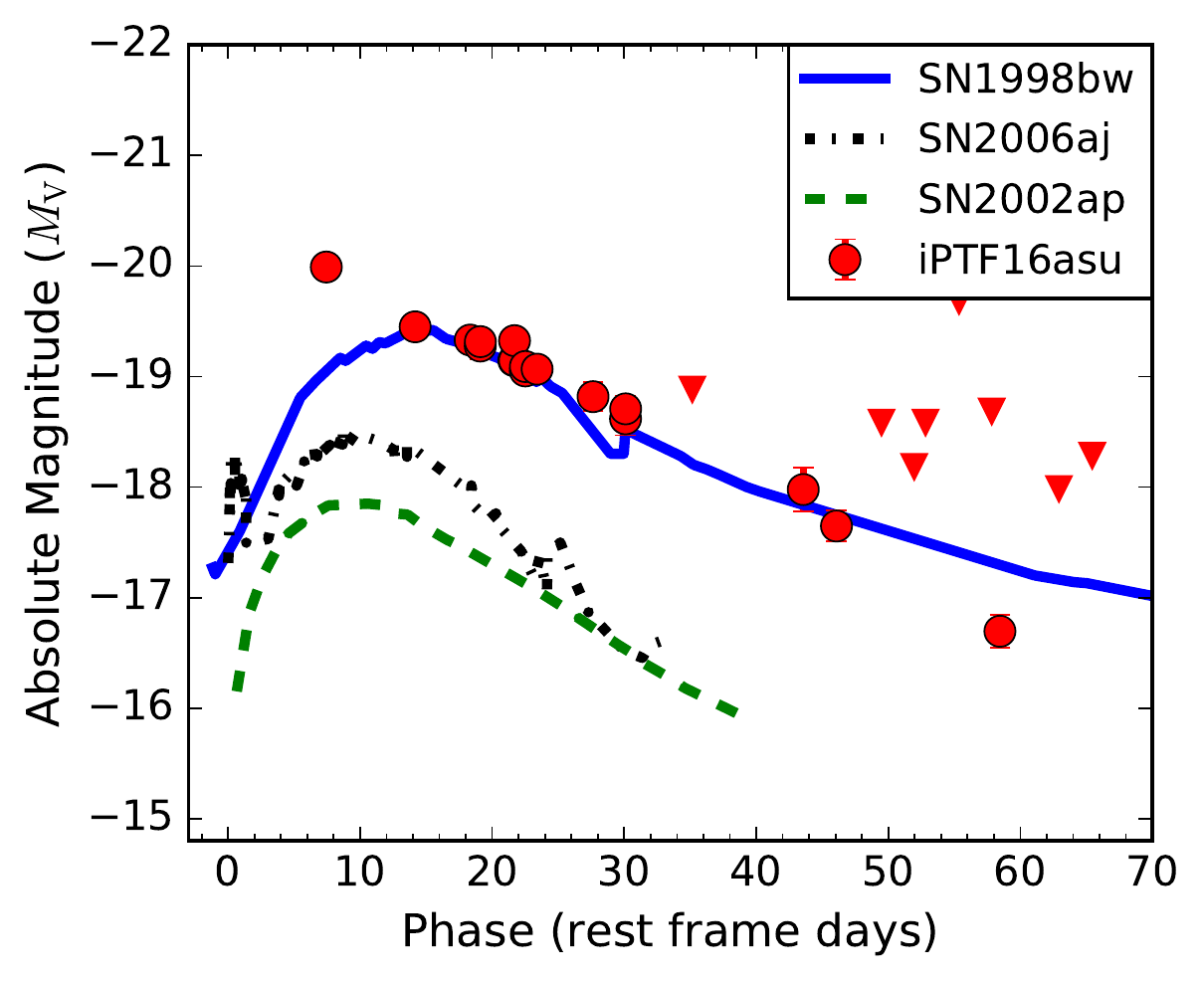}
\end{tabular}
\caption{\textit{Left:} Light curve of iPTF\,16asu (red) in the $g$~band, K-corrected to the $B$~band, compared to the light curve of SN\,1998bw (blue) in the $B$~band \citep{gvv+98, ms+99}, SN\,2006aj (black) in the $B$~band \citep{msg+06, bbh+14, bmh+14}, and SN\,2002ap (green) in the $B$~band \citep{fps+03}. \textit{Right:} Light curve of iPTF\,16asu in the $r$~band, K-corrected to the $V$~band, compared to the same SNe, all in the $V$~band.}
\label{fig:IcBLcomp}
\end{figure*}

Although $^{56}$Ni decay alone cannot explain the light curve of iPTF\,16asu, it may still contribute. Figure~\ref{fig:IcBLcomp} shows iPTF\,16asu's light curve compared to other SNe Ic-BL from the literature.
The light curve of SN\,1998bw in the $g$ and $r$~bands is a good match to that of iPTF\,16asu from $\sim15$ to 40~days, which interestingly also corresponds to the time when their spectra are very similar (Figure~\ref{fig:latespec}), suggesting that the light curve of iPTF\,16asu could plausibly be dominated by a normal SN component at these times. Their late-time slopes deviate, mainly constrained by our last $r$~band point at 60~days (Figure~\ref{fig:IcBLcomp}) -- however, the decay rates of stripped-envelope SNe are heterogeneous, and could be explained by differences in opacity and/or asymmetry, affecting the degree of gamma-ray trapping \citep{wjc15,dhy+17}. Fainter SNe Ic-BL such as SN\,2006aj and SN\,2002ap are below the light curve of iPTF\,16asu at all times (Figure~\ref{fig:IcBLcomp}). Since the light curve shows only a single, smooth peak, any $^{56}$Ni decay contribution to the total luminosity must be sub-dominant to whatever is powering the main peak.

Finally, we note that other radioactive species, such as $^{48}$Cr and $^{52}$Fe, has been proposed to power a class of fast-and-faint thermonuclear transients from He-shell detonations, so-called ``.Ia'' SNe \citep{bsw+07,skw+10}. However, given that iPTF\,16asu is 3-5~magnitudes brighter than these models predict, the spectrum at peak is blue and featureless without the expected strong \ion{Ti}{2} features, and the late-time spectrum is an excellent match to SNe Ic-BL suggesting a core-collapse explosion, we do not consider these models relevant for iPTF\,16asu.

\subsection{Magnetar}

During the core collapse of a massive star, a highly magnetized ($B \approx 10^{14}-10^{15}$~G), rapidly spinning neutron star called a magnetar can be formed. As the newborn magnetar spins down, rotational energy is released, and can significantly boost the luminosity of the SN if the spin-down time of the magnetar is comparable to the diffusion time through the ejecta \citep[e.g.,][]{kb10,woo10,mmk+15}. Magnetar models have been suggested to explain highly luminous transients, including many SLSNe as well as SN\,2011kl \citep{gmk+15,bbo+16}. iPTF\,16asu has a similar luminosity to SN\,2011kl and some relatively low luminosity SLSNe (Figures~\ref{fig:intro}, \ref{fig:lc_comp}), and so we examine whether a magnetar model is able to explain the peculiar light curve of iPTF\,16asu.

As described in \citet{kb10}, the hydrodynamic simulations for their magnetar model makes the simplifying assumption that all of the injected energy is thermalized spherically at the base of the ejecta (ignoring the possibility of anisotropic jet-like injection). They further assume homologous expansion, a shallow power law structure for interior density, and that radiation pressure dominates. An expanding bubble with a thin shell of swept up ejecta and a low density interior is formed due to central overpressure in the SN remnant, but rarely affects the outer layers of the SN ejecta. At late times the energy injected by the magnetar continues to heat the ejecta, as in $^{56}$Ni decay, but is no longer dynamically important. This process significantly affects the SN light curve.

\begin{figure}
\centering
\includegraphics[width=3.5in]{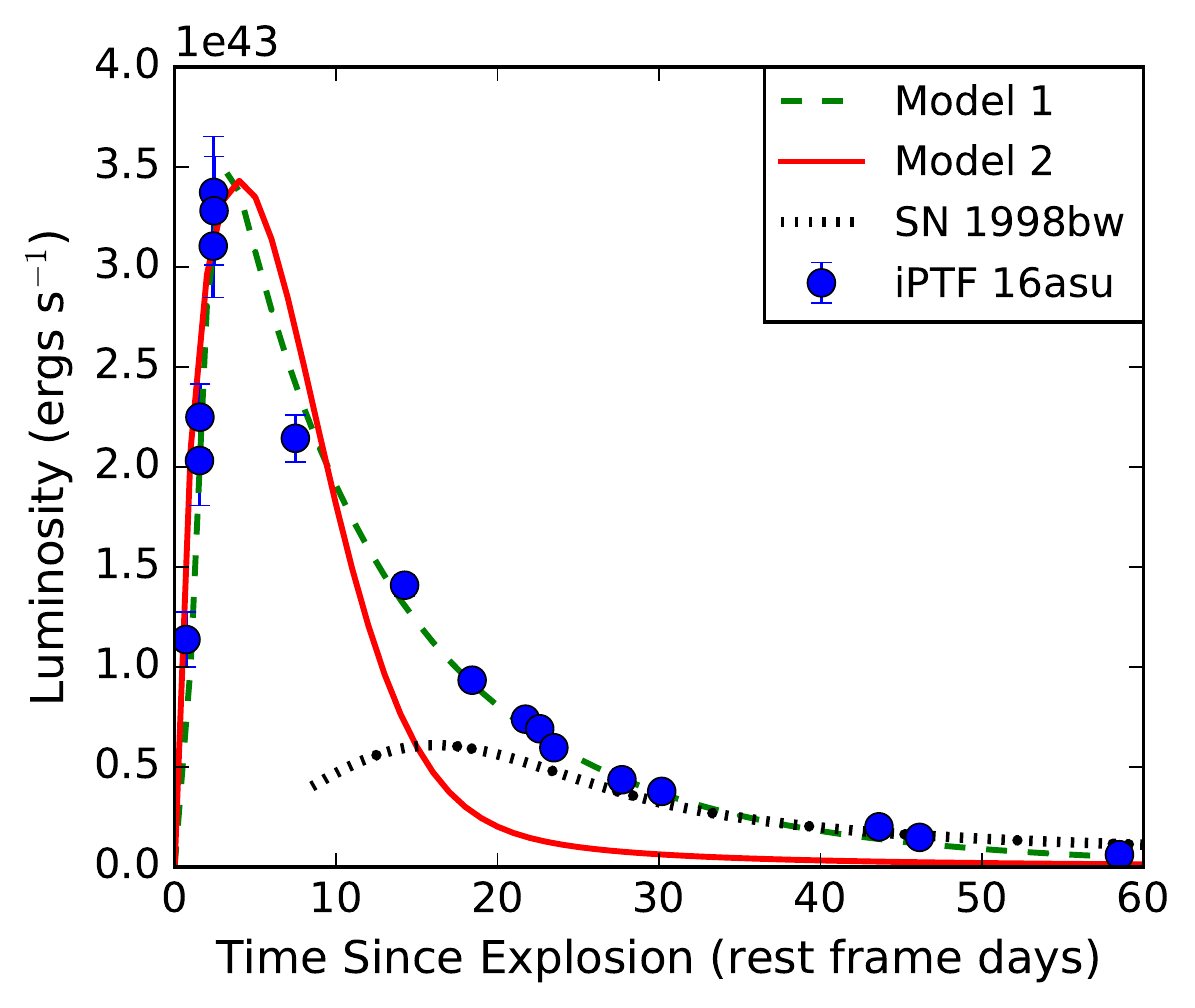}
\caption{Model 1, magnetar model best fit to the full light curve, shown in dashed green. Parameters are P$=9.95$~ms, B$=3.15 \times 10^{14}$~Gauss, and $\tau_{\rm diff}=1.8$~days. Model 2, magnetar model best fit with constrained $M_{\rm ej}=1 $M$_{\odot}$ shown as a red line. Parameters are P$=6.0$~ms, B$=4.4 \times 10^{15}$~Gauss, and $\tau_{\rm diff}=8.7$~days. The pseudobolometric light curve using $BV(RI)_{\rm c}$~bands of SN\,1998bw \citep{csc+11} is plotted in dotted black to demonstrate how $^{56}$Ni decay may power the late-time light curve.
\label{fig:magnetar}}
\end{figure}

The shape of the light curve in magnetar models depends on  three parameters: P, the initial spin period; B, the strength of the magnetic field; and $\tau_{\rm diff}$, the diffusion timescale which is proportional to M$_{\rm ej}^{1/2}$. Using the magnetar model fitting code from \citet{kbm+16} we recover the parameters B$=(3.25\pm 0.44)\times10^{14}$~Gauss, P$=10.40 \pm 0.62$~ms, and $\tau_{\rm diff} = 1.59 \pm 0.06$~days. Manually tweaking the parameters slightly to obtain a better visual fit, we show the resulting fit to the bolometric light curve in Figure~\ref{fig:magnetar} with parameters P$=9.95$~ms, B$=3.15 \times 10^{14}$~G, and $\tau_{\rm diff}=1.8$~days, and assuming an opacity $\kappa=0.1$~cm$^{2}$~g$^{-1}$. As done in the $^{56}$Ni model, the diffusion time and average velocity from the blackbody fits are used to calculate an ejecta mass of M$_{\rm ej}=0.09~$M$_{\odot}$. The parameters allow for the energy and the timescale to essentially be tuned separately, making the magnetar model quite flexible and generating a tight fit to both the peak and decay of the bolometric light curve.

Although the magnetar model produces a light curve which fits iPTF\,16asu, the derived ejecta mass of our best fit is very low. \citet{awh+16} derived similarly small ejecta masses for their rapidly-rising SNe events, which caused them to conclude the magnetar model was unlikely, while \citet{gmk+15} concluded a magnetar was a likely explanation for SN\,2011kl despite their low derived ejecta mass. For a SN Ic-BL caused by the core collapse of a massive star, a magnetar model with such a low ejecta mass would require an extreme stripping scenario to reduce the core mass. Furthermore, the \citet{kb10} magnetar model was tuned to an ejecta mass of $M_{\rm ej}=5~$M$_{\odot}$ and it is not clear that the assumptions of this model would remain valid in this low mass regime.

Another way for a magnetar model to produce a fast timescale peak, similar to that of iPTF\,16asu, is to use a small period and a high magnetic field, thereby decreasing the spin-down time. When constraining the ejecta mass to be $M_{\rm ej}=1.0~$M$_{\odot}$, we find the best fit parameters P$=6.0$~ms, B$=4.4\times10^{15}$~Gauss, and $\tau_{\rm diff}=8.7$~days. This fit is shown as a red line in Figure~\ref{fig:magnetar}, and can also reproduce the fast rise and luminous peak of iPTF\,16asu. However, this model declines too quickly to explain the entire light curve, and so one would need a two-component model (e.g. with the late-time powered by $^{56}$Ni, as was considered by \citealt{bbo+16} for SN\,2011kl). Thus, despite the compelling light curve fit, we conclude that a magnetar model is unlikely to be the sole power source of iPTF\,16asu, but remains a candidate for powering the peak emission if the late time light curve is powered by $^{56}$Ni.

\subsection{Off-Axis GRB}
\label{sec:grb}

Long GRBs are often associated with SNe Ic-BL, though not every SN Ic-BL has an accompanying GRB (see e.g. \citealt{wb06} for a review of the GRB-SN connection). GRBs are extremely energetic, relativistic and highly beamed explosions characterized by an initial flash of gamma-rays followed by an ``afterglow'' of radiation typically seen at wavelengths ranging from the X-ray to the radio. iPTF\,16asu's spectra and velocities are similar to those of SNe Ic-BL associated with GRBs (Section~\ref{sec:spec_analysis}, Figures~\ref{fig:midspec} and \ref{fig:latespec}), and so we examine whether the excess blue emission at peak could be explained as a GRB afterglow.

\begin{figure*}
\centering
\begin{tabular}{cc}
\includegraphics[width=3.5in]{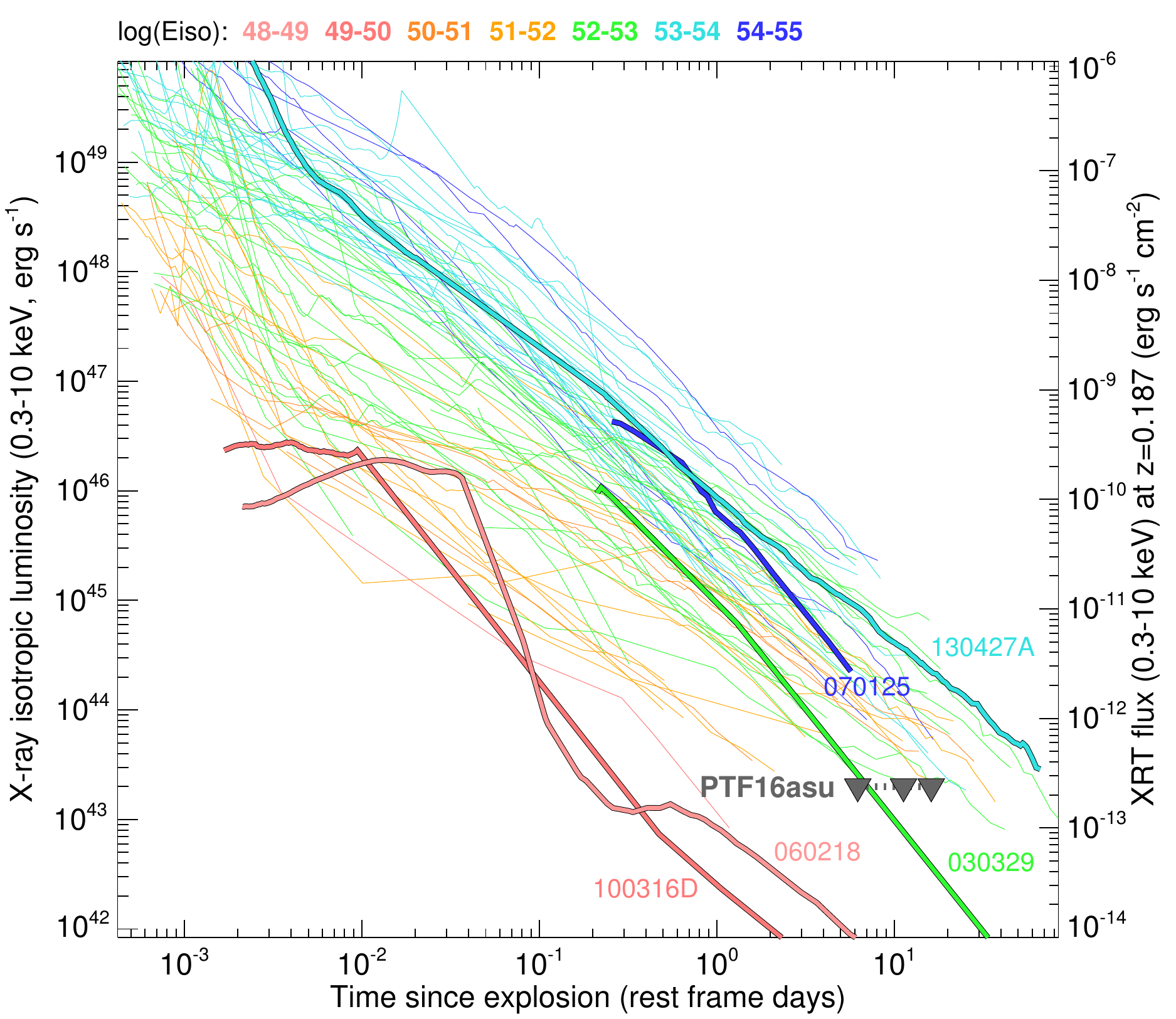} &
\includegraphics[width=3.5in]{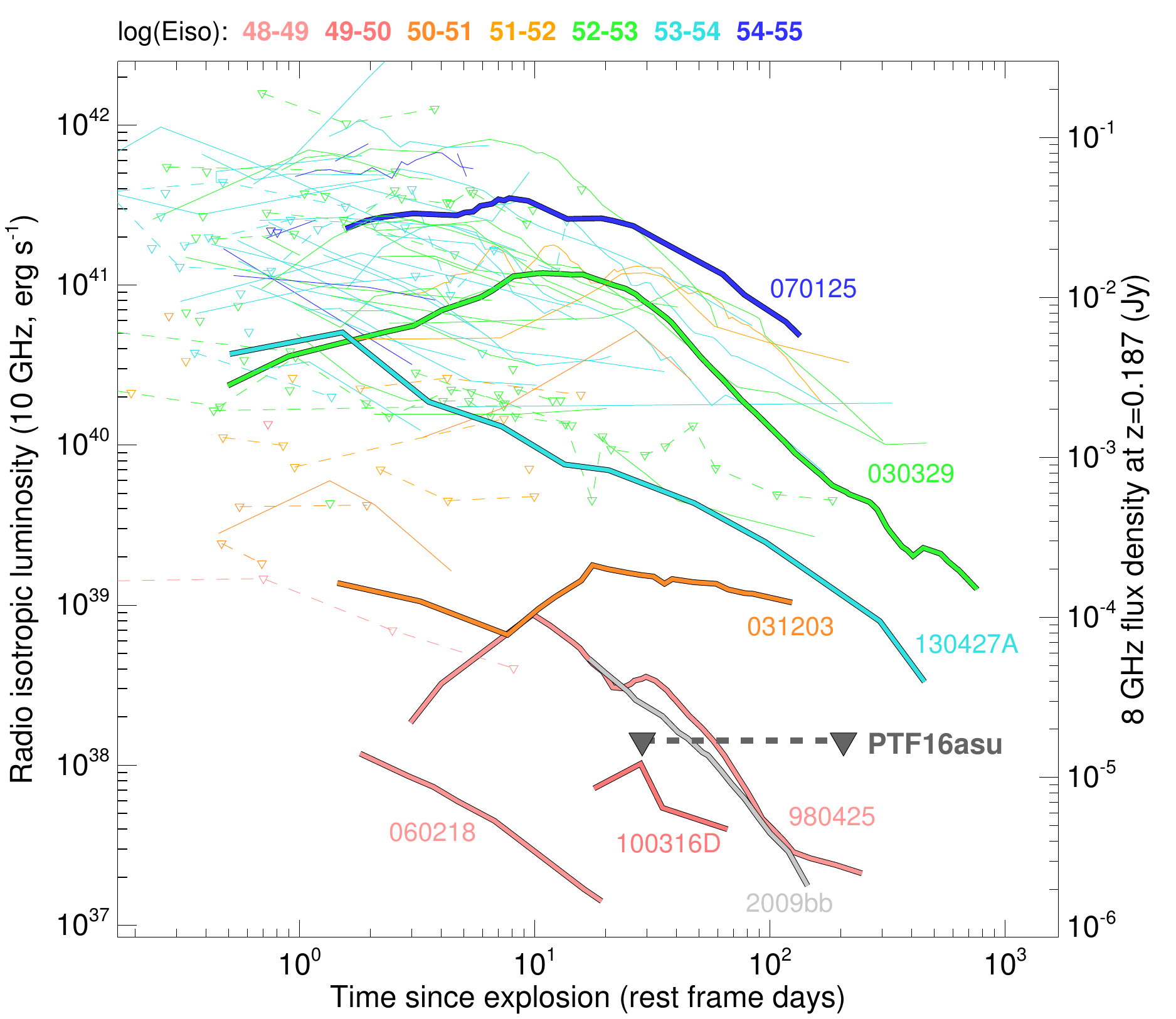}
\end{tabular}
\caption{X-ray (left) and radio (right) light curves of long-duration gamma-ray bursts, shifted to the redshift of iPTF\,16asu and plotted alongside our measured upper limits for that event from Swift and the VLA. X-ray light curves are from the Swift XRT online database \citep{ebp+07} plus \citet{tmg+04}; radio light curves are from the compilation of \citet{cf12} plus the relativistic SN\,2009bb \citep{scp+10}. Light curves are color-coded by the measured isotropic-equivalent gamma-ray luminosity of the prompt emission; see \citet{pcc+14} for additional details.  Dashed lines (on the radio plot) indicate upper limits and some prominent events are highlighted.  The upper limits on iPTF\,16asu rule out most of the parameter space for previously-observed GRB afterglows but permit a faint event similar to GRB\,060218.}
\label{fig:xray+radio}
\end{figure*}

Non-detections of iPTF\,16asu in the X-ray and radio strongly constrain the allowable GRB parameter space. The upper limits from 3 epochs of \textit{Swift} data are shown in the left-hand panel of Figure~\ref{fig:xray+radio}. While data at earlier times would have been more constraining, the upper limits rule out the bulk of observed X-ray afterglows with $E_{iso}>10^{52}$~ergs; however, weak or off-axis GRBs are not excluded by the X-ray data alone. Similarly, the right-hand panel shows the upper limits from our two epochs of VLA data. As evident from this figure, we can exclude a radio counterpart to iPTF\,16asu as luminous as SN\,1998bw or SN\,2009bb, but we cannot exclude a lower-luminosity and/or faster-evolving radio counterpart such as SN\,2006aj and SN\,2010bh. If iPTF\,16asu is associated with a GRB, then these limits suggest that it must be a faint ($E_{\rm iso}<10^{50}$~ergs) event. These constraints are consistent with the analysis from all-sky gamma-ray monitors (Section~\ref{sec:konus}), as an on-axis burst at the distance of iPTF\,16asu with ($E_{\rm iso}> 10^{50}$~ergs) would have been seen by KW or SPI-ACS.

The most unusual characteristic of iPTF\,16asu is its abrupt 4-day rise time in the optical. Such a rise time is extremely short in a SN context, but would be unprecedentedly \textit{long} in a SN-GRB context, even though optical afterglow light curves do sometimes show a rise (e.g. GRB\,970508; \citealt{ggv+98}). To explain the shape of the optical light curve as a GRB afterglow, we therefore consider off-axis GRB models.

From the NYU Afterglow Library dataset of off-axis long GRBs at an observed wavelength of 3000~\AA ($10^{15}$~Hz), the models can reproduce a 3 to 6 day rise for an observer angle between $23$ and $17$ degrees, respectively \citep{vzm10}. The dataset assumes a jet energy of $2\times10^{51}$~ergs, a jet half opening angle of $11.5$~degrees, and a homogeneous circumburst number density of $1$~cm$^{-3}$. The parameters for an observer angle of $17$~degrees produce a light curve with roughly the correct peak magnitude as iPTF\,16asu; however, changing to an observed wavelength of 30~mm (10~GHz), these parameters produce a radio light curve orders of magnitude brighter than our radio limits. Similarly, considering the low-energy models from \citet{vm11}, we find that parameters which satisfy the radio limits are inconsistently faint in the optical. The coarse grid of parameters used in \citet{vzm10} does not allow us to make precise comparisons to their model, but indicates that while a 4~day optical rise could be constructed, our optical light curve and radio upper limits cannot be simultaneously satisfied by current models. A more thorough exploration of energy and density parameter space than is available in these model grids is necessary to determine whether GRB models can account for both the bright optical emission and the lack of X-ray and radio emission.

A similar conclusion can be reached by comparing the observed spectral properties of iPTF\,16asu to typical GRB afterglows, which are well described by synchrotron radiation resulting in both a light curve and a spectrum consisting of several power law segments with associated indices \citep[e.g.][]{spn98}. If the featureless, blue spectra of iPTF\,16asu are due to a GRB afterglow, we expect the spectrum to follow a power law ($F_{\nu} \propto \nu^{-\beta}$), with typical values of the power-law index $\beta$ around 0.5-0.6 \citep[e.g.][]{kkz+10}. Fitting our first spectrum (at $+3$~days after explosion) with a power law, we find a best-fit index $\beta = -0.5$, i.e. $F_{\nu} \propto \nu^{+0.5}$, which is inconsistent with a GRB-like spectrum. In contrast, the spectrum is well-fit by a blackbody (Figure~\ref{fig:bb_spectra}). Similarly, if we compare our earliest X-ray upper limit to the corresponding point on the $r$~band light curve, we derive a limit on the optical to X-ray spectral index $\beta_{\rm OX} > 1.24$, whereas typical GRB afterglows show $\beta_{\rm OX} \sim 0.5-1.0$ \citep{gbb+08}. We also note that the decline of the light curve is better fit by exponential decay than by a power law (Section~\ref{sec:bol_lc}, Figure~\ref{fig:lum} (right)). 

While the properties of the luminous, blue peak of iPTF\,16asu do not seem to resemble a classical GRB afterglow (on- or off-axis), it is worth noting that low-luminosity GRBs like 060218 and 100316D showed thermal emission in addition to the weak synchrotron component \citep[e.g.][]{cmb+06,swl+11}. Thus, it is still possible that iPTF\,16asu could be a related phenomenon but with a significantly brighter thermal component. The origin of the thermal emission in low-luminosity GRBs is debated, though one possibility is that it is associated with shock breakout. We consider next whether such a model can also explain iPTF\,16asu.

\subsection{Shock Cooling}

The short timescales and blue colors of iPTF\,16asu are reminiscent of shock cooling transients, where the early light curve of a SN is powered by the cooling of the envelope following the breakout of the SN shock, usually followed by a second peak from the SN itself (e.g., SN\,1993J; \citealt{wbb+93}). Such a shock cooling phase should be present in all SNe \citep{ns10}, but both the duration and the luminosity will depend on the structure of the progenitor star. A peak in both the red and blue bands, as we see in iPTF\,16asu, is generally associated with shock breakout from extended material \citep{np14}. Shock cooling models have been considered for other rapidly evolving transients (e.g., \citealt{orn+10, dcs+14}) as well as low-luminosity GRBs (e.g., \citealt{nak15}), so we consider here whether iPTF\,16asu could be explained by a shock cooling scenario.

Since the peak is seen in all bands, we consider the extended envelope model of \citet{np14}. Here, the mass in the extended envelope scales as $M_e \propto \kappa^{-1} v t_{\rm peak}^2$, and the effective radius of the material scales as $R_e \propto \kappa L_{\rm peak} v^{-2}$. For the rise time and peak luminosity measured for iPTF\,16asu, this suggests an envelope mass around $\sim$0.5~M$_{\odot}$ and a lower limit on the effective radius of the material of $\sim 2\times 10^{12}$~cm, still assuming $\kappa=0.1$~cm$^{2}$~g$^{-1}$. 

\begin{figure}
\centering
\includegraphics[width=3.5in]{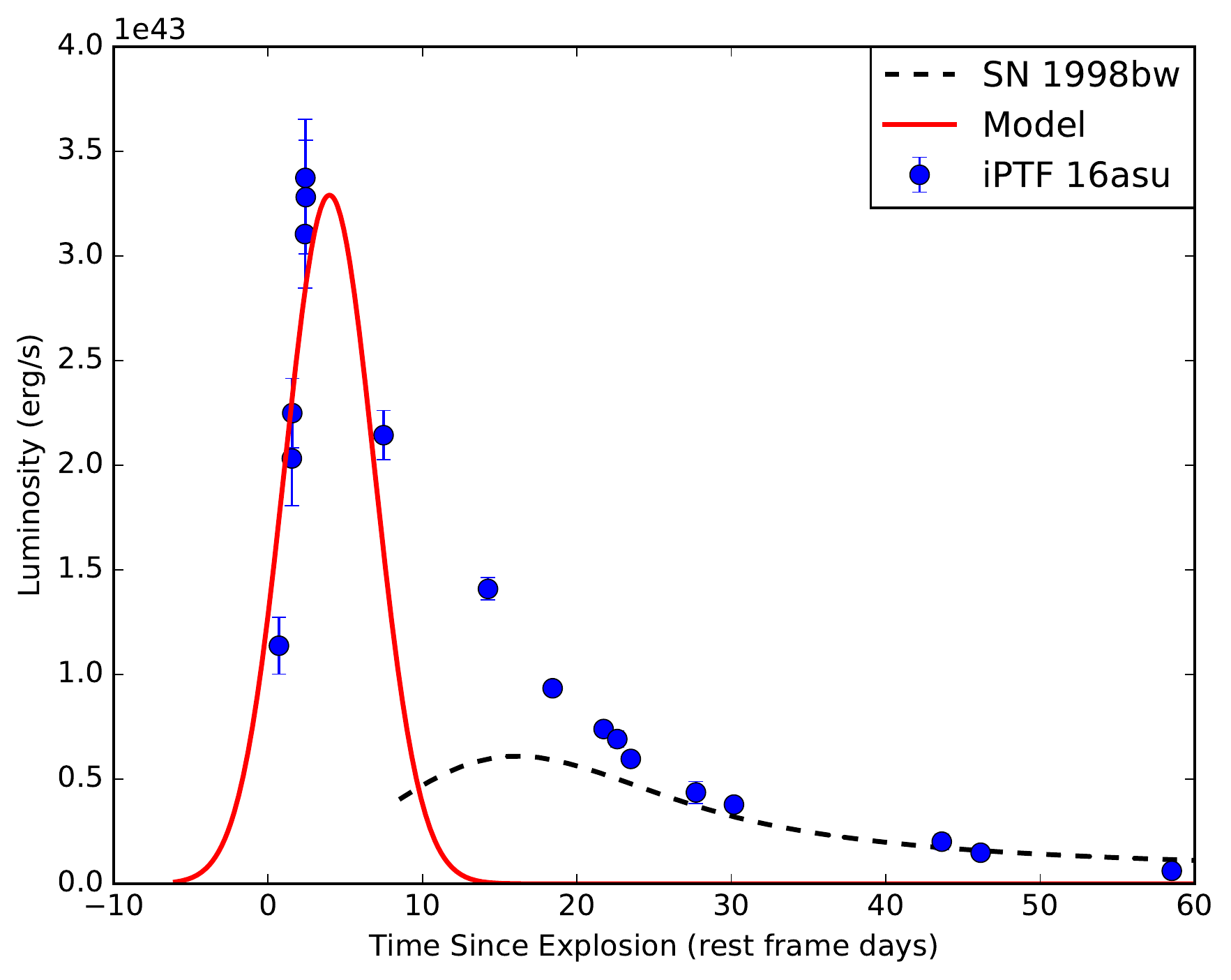}
\caption{Fit of shock-breakout model (red) from \citet{pir15} using parameters $\kappa=0.1$~cm$^{2}$~g$^{-1}$, $M_{\rm e}=0.45~$M$_{\odot}$ $R_{\rm e}=1.7\times10^{12}$~cm ($\sim 25 R_{\odot}$), and $E_e=3.8\times10^{51}$~ergs. The pseudobolometric light curve using $BV(RI)_{\rm c}$~bands of SN\,1998bw \citep{csc+11} is plotted in black to illustrate how $^{56}$Ni decay could power the late-time light curve.
\label{fig:shock}}
\end{figure}

\citet{pir15} developed this extended envelope further, and Figure~\ref{fig:shock} shows a fit of the model with the parameters $M_{\rm ej}=0.45~$M$_{\odot}$, $R_{\rm e}=1.7\times10^{12}$~cm, and $E=3.8\times10^{51}$~ergs. In general, there is a degeneracy between the initial radius of the material and the energy deposited by the shock, but our high observed velocities suggest we are in the regime of a smaller radius and higher energy \citep{pir15}. Since the energy deposited into the extended material is just a fraction of the total SN energy, if this model is correct it would imply a very high explosion energy, likely requiring a central engine.

Unlike many shock-breakout SNe, iPTF\,16asu exhibits only a single peak, so if the main light curve peak is powered by shock cooling, it must completely dominate the contribution from the underlying, normal SN light curve. The model shown in Figure~\ref{fig:shock} approximates the shock cooling light curve with a Gaussian, and is not expected to capture the decline of the light curve, which would depend on the density structure of the material. It does, however, demonstrate that an extended envelope model can produce a peak with a rise time and luminosity compatible with iPTF\,16asu. As seen in Figure~\ref{fig:IcBLcomp}, a SN Ic-BL slightly less luminous than SN\,1998bw, could be hidden underneath a luminous shock-breakout peak forming one continuous peak by the merging of the second SN peak with the decay of the first peak. 

In \citet{nak15}, SN\,2006aj/GRB\,060218 is modeled by shock breakout from energy deposited into an extended ($>100~R_{\odot}$), low-mass ($\sim0.01~$M$_{\odot}$) envelope by a low-luminosity GRB. Thus, iPTF\,16asu could have a similar explosion mechanism but with a significantly higher-mass envelope, producing a longer-duration and luminous peak. The presence of circumstellar material would also be consistent with the constraints on high-energy emission; indeed, it has been suggested that low-energy, soft GRBs like 060218 and 100316D have extended circumstellar material \citep{mgl+15}.

\section{Summary}
\label{sec:conc}
We present photometric and spectroscopic observations of the unique transient iPTF\,16asu. The key observed properties can be summarized as follows:

\begin{itemize}
    \item A rapidly evolving and luminous light curve, with a rise of 4.0~days to a high peak luminosity of $3.4\times10^{43}$~ergs~s$^{-1}$.  The decline is similarly fast, and is well fit by exponential decay with a characteristic timescale of 14 days.
    \item A blue and featureless spectrum near peak, that is well fit by a blackbody, using UV and optical data, with a temperature of $\sim 11,000~{\rm K}$ and radius of $\sim 2.5\times10^{15}$~cm.
    \item Broad spectroscopic features emerging on the decline, that are well matched to SNe Ic-BL. The velocities, as measured from the \ion{Fe}{2}~$\lambda5169$\,\AA~ line, are comparable to SNe Ic-BL with accompanying GRBs. 
    \item Non-detections in the X-ray (Swift/XRT), corresponding to limits of $(1-2)\times10^{43}$~erg~s$^{-1}$ and in the radio (VLA), corresponding to limits of $(1-2)\times10^{28}$~erg~s$^{-1}$. Non-detections by all-sky gamma-ray monitors similarly constrain any associated on-axis GRB to be low-energy ($E_{\rm iso} < 10^{50}~{\rm erg}$).
    \item A dwarf host galaxy, with a stellar mass of $\sim 5\times 10^8~{\rm M}_{\odot}$, a metallicity $Z \sim 0.3~Z_{\odot}$, and a star formation rate of $\sim 0.7~{\rm M}_{\odot}/{\rm yr}$.   
\end{itemize}

We discuss various energy sources to explain the above observed properties. We find that $^{56}$Ni decay, as in an ordinary SN Ic-BL, is adequate to explain the late time photometry. It is also consistent with the observed spectra and non-detections in the X-ray and radio bands. However, attempting to fit the rapid rise and luminous peak solely with $^{56}$Ni decay gives the unphysical result that $M_{\rm Ni}>M_{\rm ej}$. Hence we considered two different hypotheses to explain the early data.

First we considered a magnetar model. The magnetar model either requires a very small ejecta mass ($0.086~{\rm M}_{\odot}$) in order to fit the sharp rise or a high magnetic field ($B=4.4\times10^{15}$~G) that decreases the spin-down time. The latter would require that the late time data is explained by radioactive decay of $^{56}$Ni. 

Next we find that shock cooling can also explain the fast rise and high luminosity with a dense envelope ($M_{\rm e}=0.45~$M$_{\odot}$, $R_{\rm e}=1.7\times10^{12}$~cm) and high injected energy ($E_{\rm e}=3.8\times10^{51}$~erg). The required energetics in this model also implies an underlying central engine. Shock cooling through the envelope has been seen in the low-luminosity SN\,2006aj/GRB\,060218. Our spectra and kinematics are also more similar to SNe Ic-BL associated with GRBs. Our radio and X-ray limits constrain the energy ($E_{\rm iso}$) of any associated GRB to be $<10^{50}~{\rm erg}$. Regardless of whether or not there was a GRB, the late time light curve is reasonably fit by $^{56}$Ni decay.

Both of the above scenarios suggest that iPTF\,16asu was an engine-driven supernova, making it an intriguing transition object between SLSNe, low-luminosity GRBs, SNe Ic-BL, and objects like SN\,2011kl. We hope that new discoveries from the next generation of wide-field surveys (e.g. Zwicky Transient Facility; \citealt{bk+15}), will enable us to find more objects like iPTF\,16asu and more conclusively determine the origins of such fast and luminous transients.

\acknowledgements
We thank the anonymous referee for constructive comments that improved the manuscript. We thank Iair Arcavi, Maria Drout, Avishay Gal-Yam, Raffaella Margutti, and Maryam Modjaz for helpful discussions and comments. R.L. acknowledges helpful discussions at the MIAPP workshop ``Superluminous Supernovae in the Next Decade'', supported by the Munich Institute for Astro- and Particle Physics (MIAPP) of the DFG cluster of excellence ``Origin and Structure of the Universe''. We thank Harish Vedantham, Vikram Ravi and Anna Ho for assisting with the observations presented in this paper. The Intermediate Palomar Transient Factory project is a scientific collaboration among the California Institute of Technology, Los Alamos National Laboratory, the University of Wisconsin, Milwaukee, the Oskar Klein Center, the Weizmann Institute of Science, the TANGO Program of the University System of Taiwan, and the Kavli Institute for the Physics and Mathematics of the Universe. This work was supported by the GROWTH project funded by the National Science Foundation under Grant No 1545949. This work is partially based on data acquired with the Swift GO program 1215281 (grant NNX16AN84G, PI Lunnan). Part of this research was carried out at the Jet Propulsion Laboratory, California Institute of Technology, under a contract with the National Aeronautics and Space Administration. The National Radio Astronomy Observatory is a facility of the National Science Foundation operated under cooperative agreement by Associated Universities, Inc. A.C. acknowledges support from the National Science Foundation CAREER award n. 1455090. D.S. and D.F. gratefully acknowledge support from RSF grant 17-12-01378. This work made use of the data products generated by the NYU SN group, and released under DOI:10.5281/zenodo.58767, available at \url{https://github.com/nyusngroup/SESNspectraLib}. Based on observations made with the Nordic Optical Telescope, operated by the Nordic Optical Telescope Scientific Association at the Observatorio del Roque de los Muchachos, La Palma, Spain, of the Instituto de Astrofisica de Canarias. Some of the data presented herein were obtained at the W.M. Keck Observatory, which is operated as a scientific partnership among the California Institute of Technology, the University of California and the National Aeronautics and Space Administration. The Observatory was made possible by the generous financial support of the W.M. Keck Foundation. The authors wish to recognize and acknowledge the very significant cultural role and reverence that the summit of Mauna Kea has always had within the indigenous Hawaiian community.  We are most fortunate to have the opportunity to conduct observations from this mountain.

\textit{Facilities:} \facility{PO:1.2m}, \facility{PO:1.5m}, \facility{PO:Hale}, \facility{Keck:I}, \facility{Keck:II}, \facility{NOT}, \facility{TNG}, \facility{Swift}, \facility{VLA}

\clearpage
\LongTables

\begin{deluxetable}{lcccccccc}
\tablecaption{Log of iPTF\,16asu Photometric Observations}
\tablehead{
\colhead{Observation Date} &
\colhead{Phase\tablenotemark{a}}  &
\colhead{Filter} &
\colhead{Magnitude\tablenotemark{b}} &
\colhead{Telescope}  &\\
\colhead{(MJD)} &
\colhead{(rest-frame days)} &
\colhead{} &
\colhead{(AB)} &
\colhead{} &
}
\startdata
57508.32 & -12.50 & g & $>20.61$ & P48 \\
57510.27 & -10.87 & g & $>20.89$ & P48 \\
57510.30 & -10.84 & g & $>20.80$ & P48 \\
57510.33 & -10.82 & g & $>20.91$ & P48 \\
57511.26 & -10.03 & g & $>20.78$ & P48 \\
57511.29 & -10.00 & g & $>20.72$ & P48 \\
57511.32 & -9.98 & g & $>20.53$ & P48 \\
57512.26 & -9.18 & g & $>21.05$ & P48 \\
57512.29 & -9.16 & g & $>20.82$ & P48 \\
57512.32 & -9.14 & g & $>21.09$ & P48 \\
57513.25 & -8.35 & g & $>20.96$ & P48 \\
57513.28 & -8.33 & g & $>20.96$ & P48 \\
57513.31 & -8.30 & g & $>20.72$ & P48 \\
57519.26 & -3.29 & g & 20.43 $\pm$ 0.13 & P48 \\
57519.29 & -3.27 & g & $>20.29$ & P48 \\
57519.32 & -3.24 & g & $>20.01$ & P48 \\
57520.25 & -2.46 & g & 19.80 $\pm$ 0.12 & P48 \\
57520.28 & -2.43 & g & 19.69 $\pm$ 0.08 & P48 \\
57521.26 & -1.61 & g & 19.34 $\pm$ 0.09 & P48 \\
57521.29 & -1.58 & g & 19.25 $\pm$ 0.09 & P48 \\
57521.32 & -1.56 & g & 19.28 $\pm$ 0.09 & P48 \\
57525.40 & 1.88 & g & 19.38 $\pm$ 0.09 & P60 \\
57527.34 & 3.51 & g & 19.51 $\pm$ 0.07 & P60 \\
57535.33 & 10.24 & g & 20.43 $\pm$ 0.09 & P60 \\
57538.34 & 12.78 & g & 20.87 $\pm$ 0.05 & P60 \\
57540.30 & 14.42 & g & 21.01 $\pm$ 0.07 & P60 \\
57544.21 & 17.72 & g & 21.49 $\pm$ 0.08 & P60 \\
57545.25 & 18.60 & g & 21.48 $\pm$ 0.11 & P60 \\
57545.26 & 18.61 & g & 21.14 $\pm$ 0.09 & P60 \\
57546.31 & 19.49 & g & 21.69 $\pm$ 0.13 & P60 \\
57551.35 & 23.73 & g & $>22.09$ & P60 \\
57560.26 & 31.24 & g & $>20.29$ & P60 \\
57580.20 & 48.03 & g & $>21.69$ & P60 \\
57581.23 & 48.90 & g & $>21.19$ & P60 \\
57584.24 & 51.43 & g & $>21.49$ & P60 \\
57587.18 & 53.91 & g & $>21.69$ & P60 \\
57587.88 & 54.50 & g & $>23.09$ & TNG \\ 
57527.33 & 3.51 & r & 19.60 $\pm$ 0.09 & P60 \\
57535.32 & 10.23 & r & 20.19 $\pm$ 0.07 & P60 \\
57540.27 & 14.40 & r & 20.34 $\pm$ 0.04 & P60 \\
57541.18 & 15.17 & r & 20.40 $\pm$ 0.12 & P60 \\
57541.18 & 15.17 & r & 20.36 $\pm$ 0.07 & P60 \\
57544.19 & 17.70 & r & 20.55 $\pm$ 0.05 & P60 \\
57544.25 & 17.75 & r & 20.55 $\pm$ 0.04 & P60 \\
57544.26 & 17.76 & r & 20.37 $\pm$ 0.03 & P60 \\
57545.23 & 18.58 & r & 20.65 $\pm$ 0.07 & P60 \\
57545.23 & 18.58 & r & 20.61 $\pm$ 0.09 & P60 \\
57546.29 & 19.47 & r & 20.64 $\pm$ 0.05 & P60 \\
57551.32 & 23.71 & r & 20.89 $\pm$ 0.13 & P60 \\
57554.24 & 26.17 & r & 21.09 $\pm$ 0.15 & P60 \\
57554.25 & 26.17 & r & 21.00 $\pm$ 0.12 & P60 \\
57560.24 & 31.22 & r & $>20.83$ & P60 \\
57570.22 & 39.62 & r & 21.73 $\pm$ 0.20 & P60 \\
57573.21 & 42.14 & r & 22.06 $\pm$ 0.14 & P60 \\
57577.25 & 45.55 & r & $>21.13$ & P60 \\
57580.19 & 48.02 & r & $>21.53$ & P60 \\
57581.22 & 48.89 & r & $>21.13$ & P60 \\
57584.23 & 51.42 & r & $>20.03$ & P60 \\
57587.17 & 53.90 & r & $>21.03$ & P60 \\
57587.90 & 54.52 & r & 23.01 $\pm$ 0.15 & TNG \\
57593.21 & 58.99 & r & $>21.73$ & P60 \\
57596.21 & 61.51 & r & $>21.43$ & P60 \\
57525.40 & 1.88 & i & 19.57 $\pm$ 0.14 & P60 \\
57527.33 & 3.51 & i & 19.66 $\pm$ 0.07 & P60 \\ 
57535.33 & 10.24 & i & 19.87 $\pm$ 0.05 & P60 \\
57538.33 & 12.77 & i & 20.09 $\pm$ 0.05 & P60 \\
57540.28 & 14.41 & i & 20.29 $\pm$ 0.06 & P60 \\
57544.20 & 17.71 & i & 20.46 $\pm$ 0.05 & P60 \\
57544.21 & 17.72 & i & 20.43 $\pm$ 0.05 & P60 \\
57544.26 & 17.77 & i & 20.43 $\pm$ 0.06 & P60 \\
57545.24 & 18.59 & i & 20.48 $\pm$ 0.11 & P60 \\
57545.25 & 18.59 & i & 20.46 $\pm$ 0.10 & P60 \\
57546.29 & 19.48 & i & 20.68 $\pm$ 0.08 & P60 \\
57551.33 & 23.72 & i & $>20.84$ & P60 \\
57554.25 & 26.18 & i & 20.82 $\pm$ 0.16 & P60 \\
57554.26 & 26.19 & i & 20.73 $\pm$ 0.12 & P60 \\
57560.25 & 31.23 & i & $>20.55$ & P60 \\
57570.23 & 39.63 & i & $>21.25$ & P60 \\
57573.21 & 42.15 & i & $>21.75$ & P60 \\
57580.20 & 48.03 & i & $>20.64$ & P60 \\
57581.22 & 48.89 & i & $>21.14$ & P60 \\
57584.23 & 51.43 & i & $>20.05$ & P60 \\
57584.26 & 51.45 & i & $>20.75$ & P60 \\
57587.17 & 53.90 & i & $>21.25$ & P60 \\
57587.89 & 54.51 & i & 23.07 $\pm$ 0.16 & TNG \\
57593.22 & 58.99 & i & $>20.95$ & P60 \\
57596.21 & 61.51 & i & $>21.14$ & P60 \\
57527.29 & 3.47 & V & $>19.48$ & Swift \\
57527.29 & 3.47 & B & 19.45 $\pm$ 0.2 & Swift \\
57527.29 & 3.47 & u & 19.6 $\pm$ 0.14 & Swift \\
57527.29 & 3.47 & UVW1 & 20.52 $\pm$ 0.14 & Swift \\ 
57527.29 & 3.47 & UVW2 & 21.8 $\pm$ 0.19 & Swift \\
57527.29 & 3.47 & UVM2 & 21.27 $\pm$ 0.14 & Swift \\
57534.33 & 9.4 & V & $>18.95$ & Swift \\
57534.33 & 9.4 & B & $>19.61$ & Swift \\
57534.33 & 9.4 & u & $>20.37$ & Swift \\
57534.33 & 9.4 & UVW1 & $>21.46$ & Swift \\
57534.33 & 9.4 & UVW2 & $>22.46$ & Swift \\
57534.33 & 9.4 & UVM2 & $>22.48$ & Swift \\
57541.19 & 9.4 & V & $> 18.96$ & Swift \\
57541.19 & 9.4 & B & $> 19.86$ & Swift \\
57541.19 & 9.4 & u & $> 20.68$ & Swift \\
57541.19 & 9.4 & UVW1 & $> 21.78$ & Swift \\
57541.19 & 9.4 & UVW2 & $> 22.60$  & Swift \\
57541.19 & 9.4 & UVM2 & $> 22.56$ & Swift 
\enddata
\label{tab:phot}
\tablenotetext{a}{Phase is in rest-frame days relative to bolometric maximum light.}
\tablenotetext{b}{Corrected for Galactic extinction.}
\end{deluxetable}

\begin{deluxetable}{lcccccccc}
\tablecaption{Log of iPTF\,16asu Spectroscopic Observations}
\tablehead{
\colhead{Observation Date} &
\colhead{Phase\tablenotemark{a}}  &
\colhead{Instrument} &
\colhead{Grating} &
\colhead{Filter}  &
\colhead{Wavelength} &
\colhead{Resolution} &
\colhead{Exp. Time} &
\colhead{Airmass} \\
\colhead{} &
\colhead{(rest-frame days)} &
\colhead{} &
\colhead{} &
\colhead{} &
\colhead{(\AA)} &
\colhead{(\AA)} &
\colhead{(s)} &
\colhead{}
}
\startdata
2016 May 14.30 & $-0.73$ & P200+DBSP & 600/4000   & None  & 3101$-$9199 & 1.30 & 600 & 1.21 \\
2016 May 16.06   & $+0.75$ & NOT+ALFOSC  & GRISM 4    & None & 3478$-$9662 & 3.35 & 2400 & 1.35 \\
2016 May 24.97   & $+8.25$ & TNG+DOLORES  & LR-B + LR-R   & None & 3315$-$10330 & 2.65 & 2100 & 1.09 \\
2016 May 27.36   & $+10.27$ & P200+DBSP  & 600/4000   & None & 3600$-$10237 & 1.30 & 1800 & 1.69 \\
2016 Jun 04.39   & $ +17.03 $ & Keck2+DEIMOS & 600ZD  & GG455  & 4550$-$9649 & 0.65 & 1000 & 1.29 \\
2016 Jun 07.36   & $+19.53$ & Keck1+LRIS     & 400/3400, 400/8500  & None  & 3072$-$10285 & 1.55 & 950 & 1.17 \\
2016 Jun 10.42   & $+22.11$ & Keck1+LRIS     & 400/3400, 400/8500  & None  & 3101$-$10290 & 1.55 & 980 & 1.71 \\
2016 Jul 06.30   & $+43.92$ & Keck1+LRIS   & 400/3400, 400/8500  & None & 3067$-$10289 & 1.55 & 2400 & 1.30 \\
2017 Apr 29.39 & $+294.04$ & Keck1+LRIS & 400/3400, 400/8500  & None & 3063$-$10318 & 1.55 & 2400 & 1.02 
\enddata
\label{tab:spec}
\tablenotetext{a}{Phase is in rest-frame days relative to bolometric maximum light (MJD 57523.25).}
\end{deluxetable}

\begin{deluxetable}{ccccccccc}
\tablecaption{$K$-corrections\tablenotemark{a} Derived from Spectra}
\tablehead{
\colhead{Phase\tablenotemark{b}} &
\colhead{$K_{\rm rr}$}  &    
\colhead{$K_{\rm gr}$}  &
\colhead{$K_{\rm gg}$}  &
\colhead{$K_{\rm ug}$}  &
\colhead{$K_{\rm ii}$}  &
\colhead{$K_{\rm ri}$}  &
\colhead{$K_{\rm Bg}$}  &
\colhead{$K_{\rm Vr}$} \\
\colhead{(rest-frame days)} &
\colhead{(mag)}  &
\colhead{(mag)}  &
\colhead{(mag)}  &
\colhead{(mag)}  &
\colhead{(mag)}  &
\colhead{(mag)}  &
\colhead{(mag)}  &
\colhead{(mag)} 
}
\startdata
$-0.73$  &   -0.368  &  -0.041 &  -0.270  & -0.194   &   \nodata  &  -0.173  & -0.192  & -0.269 \\
$+0.75$ &   -0.357 &   -0.062 &  -0.260  & -0.215   &   \nodata  & -0.164 &  -0.177  & -0.258 \\
$+8.25$  &  -0.136  &  -0.286 &   0.056 &  -0.372  & -0.285  & -0.206 &   -0.078 &  -0.175 \\
$+10.27$  &   -0.107  &  -0.280  &  0.062 &  -0.438 &  -0.325  & -0.204 &  -0.079  & -0.158 \\
$+17.03$  &  -0.109    &   \nodata   &   \nodata   &   \nodata     & \nodata  & -0.199    &  \nodata  & -0.121 \\
$+19.53$  &  -0.043  &  -0.467  &  0.410  & -0.688  & -0.255 &  -0.195  &  0.102  & -0.112 \\
$+22.11$  &  -0.050  &  -0.487  &  0.422 &  -0.768  & -0.214  & -0.187  &  0.127  & -0.110 
\enddata
\label{tab:kcor}
\tablenotetext{a}{$K$-correction defined as in \citet{hbb+02}, so that for filters $Q$ and $R$, $m_{\rm R} = M_{\rm Q} + DM + K_{\rm QR}$.}
\tablenotetext{b}{Phase is in rest-frame days relative to bolometric maximum light (MJD 57523.25).}
\end{deluxetable}

\begin{deluxetable}{lcccccccc}
\tablecaption{Spectral Line Velocities of iPTF\,16asu}
\tablehead{
\colhead{Observation Date} &
\colhead{Phase\tablenotemark{a}}  &
\colhead{\ion{Fe}{2} Velocity} &
\colhead{\ion{Fe}{2} Broadening} &
\colhead{\ion{Si}{2} Velocity}\\
\colhead{} &
\colhead{(rest-frame days)} &
\colhead{(1000 km~s$^{-1}$)} &
\colhead{(1000 km~s$^{-1}$)} &
\colhead{(1000 km~s$^{-1}$)} &
}
\startdata
2016 May 24.97   & $+8.25$ & $28.3^{+1.1}_{-1.3}$ & $5.5^{+1.0}_{-1.2}$ & \\
2016 May 27.36   & $+10.27$ & $29.5^{+1.0}_{-1.4}$ & $5.9^{+1.0}_{-1.3}$ & 23.3\\
2016 Jun 04.39   & $+17.03$ & $25.7_{-0.3}^{+0.3}$ & $5.1^{+0.3}_{-0.3}$ & 19.8\\
2016 Jun 07.36   & $+19.53$ & $21.6_{-0.4}^{+0.4}$ & $4.4^{+0.4}_{-0.5}$ & 19.2\\
2016 Jun 10.42   & $+22.11$ & $22.0_{-1.3}^{+1.0}$ & $4.3^{+1.3}_{-1.3}$ & 16.8
\enddata
\label{tab:vel}
\tablenotetext{a}{Phase is in rest-frame days relative to bolometric maximum light (MJD 57523.25).}
\end{deluxetable}

\begin{deluxetable}{lc}
\tablecaption{Host Galaxy Emission Line Fluxes}
\tablehead{
\colhead{Line} &
\colhead{Flux} \\
\colhead{} &
\colhead{($10^{-16}~{\rm erg~s}^{-1}~{\rm cm}^{-2}$)} 
}
\startdata
{[\ion{O}{2}]} 3727  & 3.50 $\pm$ 0.10 \\
{[\ion{Ne}{3}]} 3869 & 0.62 $\pm$ 0.08 \\
H$\gamma$ 4341    & 0.56 $\pm$ 0.07 \\
H$\beta$ 4861     & 1.41 $\pm$ 0.08 \\
{[\ion{O}{3}]} 4959 & 1.91 $\pm$ 0.12 \\
{[\ion{O}{3}]} 5007 & 5.72 $\pm$ 0.09 \\
H$\alpha$ 6563    & 5.01 $\pm$ 0.09 \\
{[\ion{N}{2}]} 6583 & 0.24 $\pm$ 0.10 \\
{[\ion{S}{2}]} 6717 & 0.66 $\pm$ 0.13 \\
{[\ion{S}{2}]} 6731 & 0.41 $\pm$ 0.14
\enddata
\label{tab:lineflux}
\end{deluxetable}

\end{document}